%% file: main.tex
\documentclass[10pt,twocolumn]{z2-article}
\pdfoutput=1

\usepackage{framed}
\usepackage{times}
\usepackage{oneinchmargins}
\usepackage{setspace}
\usepackage{multirow}
\usepackage{amsmath}
\usepackage{centernot}
\usepackage{amssymb}
\usepackage[hyphens]{url}
\usepackage{alltt}
\usepackage{listings}
\usepackage{cite}
\usepackage{fancyvrb}
\usepackage{relsize}
\usepackage{soul}
\usepackage{array}
\usepackage[position=bottom]{subfig}
\usepackage{booktabs}
\usepackage{enumitem}
\usepackage{currfile}
\usepackage[usenames,dvipsnames,svgnames,table]{xcolor}
\usepackage{tikz}
\usepackage{pbox}
\usepackage{xstring}
\usepackage{xspace}
\usepackage{epstopdf}
\usepackage{graphicx}
\usepackage{float}
\usepackage{makecell}
\usepackage{url}
\usepackage{fixltx2e}

\makeatletter
\def\url@leostyle{%
  \@ifundefined{selectfont}{\def\UrlFont{\sf}}{\def\UrlFont{\small\ttfamily}}}
\makeatother
\floatstyle{boxed}
\newfloat{aside}{bth!}{lop}
\floatname{aside}{Aside}

\newcommand*\circled[1]{\tikz[baseline=(char.base)]{
            \node[shape=circle,draw,inner sep=0.5pt] (char) {#1};}}

\newlength{\tempa}

\usepackage[available,functional,reproduced]{usenixbadges}
 
\begin{document}
% \pagenumbering{gobble}
\input{macros}
\begin{spacing}{1.0}
\input{title}
\input{abstract}
\input{intro}

\input{bg}
\input{mot}
\input{llsm}

\input{eval}
\input{new.related}
\input{conc}
\input{ack}
%\clearpage
\input{bib}
\end{spacing}
\end{document}

%% file: macros.tex
% Haryadi .. parskip, and par-indent
%\setlength\parindent{0pt}
%\setlength\parskip{5pt}

\newcommand{\beforesec}{\vspace{-.4cm}}
\newcommand{\aftersec}{\vspace{-.3cm}}

\newcommand{\beforesect}{\vspace{-.1cm}}
\newcommand{\aftersect}{\vspace{-.3cm}}

\newcommand{\beforesub}{\vspace{-.3cm}}
\newcommand{\aftersub}{\vspace{-.25cm}}

\newcommand{\beforesubsub}{\vspace{-.2cm}}
\newcommand{\aftersubsub}{\vspace{-.2cm}}

 \newcommand{\zsection}[1]{\section{#1}}
 \newcommand{\zsubsection}[1]{\subsection{#1}}
 \newcommand{\zsubsubsection}[1]{\subsubsection{#1}}

\newcommand{\smush}{0.25in}

\newcommand{\figWidthOne}{3.05in} 
\newcommand{\figWidthHalf}{1.45in} 
\newcommand{\figWidthTwo}{3.05in} 
\newcommand{\figWidthTwop}{1.6in} 
\newcommand{\figWidthThree}{2.2in} 
\newcommand{\figWidthSix}{1.1in} 
\newcommand{\figWidthFour}{1.7in} 
\newcommand{\figHeight}{2.0in}
\newcommand{\captionText}[2]{\textbf{#1} \textit{\small{#2}}}

\newcommand{\zbullet}{\hspace{-0.1cm}$\bullet$}

\newcommand{\eg}{e.g.}
\newcommand{\ie}{i.e.}
\newcommand{\etal}{et al.}
\newcommand{\apriori}{\textit{a priori}}

%-------------------------------------------------------------------
% Haryadi -- new command
\newcommand{\msub}[1]{\vspace{1pt}\noindent{\bf #1}}

\newcommand{\ts}[1]{{\tt{\small#1}}}
\newcommand{\tsb}[1]{{\tt{\small{\bf#1}}}}
\newcommand{\tse}[1]{{\tt{\small{\em#1}}}}
\newcommand{\tss}[1]{{\tt{\footnotesize#1}}}
\newcommand{\exc}{$^{\ddag}$}        % except
\newcommand{\EIO}{\ts{EIO}}
\newcommand{\ENOSPC}{\ts{ENOSPC}}
\newcommand{\EDQUOT}{\ts{EDQUOT}}

%-------------------------------------------------------------------
% For fault injection table
\newcommand{\ip}{bad}
\newcommand{\nullp}{$\emptyset$}

\newcommand{\oops}{o}
\newcommand{\dead}{$\times$}
\newcommand{\alive}{$\surd$}
\newcommand{\unuse}{$\times$}
\newcommand{\use}{$\surd$}
\newcommand{\ic}{$\times$}
\newcommand{\con}{$\surd$}
\newcommand{\gpf}{G}
\newcommand{\npe}{null-pointer}
\newcommand{\usebuta}{$\surd^a$} % unmountable
\newcommand{\hwdetect}{d}

%% TABLE 2
\newcommand{\lateoops}{o$^b$}  % oops happens, but late
\newcommand{\lategpf}{G$^b$}   % gpf, but late
\newcommand{\iop}{i}           % invalid opcode
\newcommand{\detects}{d}       % assertion
\newcommand{\silentret}{s}     % app fails silently
\newcommand{\errorret}{e}      % app returns an error
\newcommand{\appworks}{$\surd$}     % app works!
\newcommand{\usebutar}{$\surd^{ar}$} % read-only and unmountable

\newcommand{\tbls}{\hspace{0.025in}}
\newcommand{\tblss}{\hspace{0.015in}}
\newcommand{\shrinkless}{\vspace{-0.01cm}}

%-------------------------------------------------------------------

%-------------------------------------------------------------------

% axes         
\newcommand{\x}{{\em x}}
\newcommand{\y}{{\em y}}
\newcommand{\xaxis}{x-axis}
\newcommand{\yaxis}{y-axis}

\newcommand{\KB}{~KB}
\newcommand{\KBs}{~KB/s}
\newcommand{\Kbs}{~Kbit/s}
\newcommand{\mbs}{~Mbit/s}
\newcommand{\MB}{~MB}
\newcommand{\GB}{~GB}
\newcommand{\MBs}{~MB/s}
\newcommand{\mus}{\mbox{$\mu s$}}
\newcommand{\ms}{\mbox{$ms$}}

\newcommand{\unix}{{\sc Unix}}

\newcommand{\bquote}{\vspace{-0.25cm} \begin{quote}}
\newcommand{\equote}{\end{quote}\vspace{-0.05cm} }

\newcommand{\zquote}[2]{\begin{quote}
#1 --
{\em``#2'' }
%{\bf -- #1 }
\end{quote}}

\newcommand{\XXX}[1]{{\small {\bf (XXX: #1)}}}

\newcommand{\XXXX}{{\bf XXX}}
\newcommand{\xx}{{\bf XXX}}

% normal
\newcommand{\beforecaption}{\begin{spacing}{0.80}}
\newcommand{\aftercaption}{\end{spacing}}
\newcommand{\mycaption}[3]{{\beforecaption\caption{\label{#1}{ \bf #2. } {\em  #3}}\aftercaption}}

\newcommand{\sref}[1]{\S\ref{#1}}

\newcommand{\xxx}[1]{  \underline{ {\small {\bf (XXX: #1)}}}}

\newenvironment{packeditemize}{
\begin{itemize}
  \setlength{\itemsep}{1pt}
  \setlength{\parskip}{0pt}
  \setlength{\parsep}{0pt}
}{\end{itemize}}

\newcommand{\smalltt}[1]{\texttt{\fontsize{8.7}{5}\selectfont #1}}
\newcommand{\llename}{\smalltt{LLE}}
\newcommand{\mapname}{\smalltt{LLE-MAP}}
\newcommand{\mapnames}{\smalltt{LLE-MAPs}}
\newcommand{\numsys}{eight}
\newcommand{\numvuls}{26}
\newcommand{\fwrong}{$\times$}
\newcommand{\smallit}[1]{\textit{\scriptsize #1}}
\newcommand{\verysmalltt}[1]{\texttt{\scriptsize #1}}
\newcommand{\verysmall}[1]{\scriptsize #1}
\newcommand*\rot{\rotatebox{90}}
\newcommand{\hardraftname}{\mbox{\textsc{\fontsize{9.2}{5}\selectfont Haft}}}
\newcommand{\giname}{\mbox{\textsc{\fontsize{9.2}{5}\selectfont Saucr}}}
\newcommand{\ginamesmall}{\mbox{\textsc{\fontsize{7.2}{5}\selectfont Saucr}}}
\newcommand{\ginamebig}{\mbox{\textsc{\fontsize{10}{6}\selectfont Saucr}}}
\newcommand{\storesys}{\mbox{\textsc{\fontsize{10}{6}\selectfont Sacs}}}
\newcommand{\parname}{\mbox{\textsc{\fontsize{9.2}{5}\selectfont Par}}}
\newcommand{\ctrlname}{\mbox{\textsc{\fontsize{9.2}{5}\selectfont Ctrl}}}
\newcommand{\rasorname}{\mbox{\textsc{\fontsize{9.2}{5}\selectfont Rasor}}}
\newcommand{\sysname}{\mbox{\textsc{\fontsize{9.2}{5}\selectfont Bourbon}}}
\newcommand{\sysnamebig}{\mbox{\textsc{\fontsize{10}{6}\selectfont Bourbon}}}
\newcommand{\sysnamesmall}{\mbox{\textsc{\fontsize{9.2}{5}\selectfont Bourbon}}}
% ctrl local storage layer
\newcommand{\ctrlstore}{\mbox{\textsc{\fontsize{9.2}{5}\selectfont Clstore}}}
\newcommand{\ctrlstoresmall}{\mbox{\textsc{\fontsize{7.2}{5}\selectfont Clstore}}}

\newcommand{\ctrlnamesmall}{\mbox{\textsc{\fontsize{7.2}{5}\selectfont Ctrl}}}
\newcommand{\ctrlnaive}{\mbox{\textsc{\fontsize{9.2}{5}\selectfont Ctrl-Naive}}} 
\newcommand{\ctrlnaivesmall}{\mbox{\textsc{\fontsize{7.2}{5}\selectfont Ctrl-Naive}}} 

\newcommand{\ctrlraft}{\mbox{\textsc{\fontsize{9.2}{5}\selectfont Ctrl-Raft}}} 
\newcommand{\ctrlzab}{\mbox{\textsc{\fontsize{9.2}{5}\selectfont Ctrl-Zab}}} 
\newcommand{\errfsname}{\mbox{\textit{\fontsize{11.2}{5}\selectfont errfs}}}
\newcommand{\errbenchname}{\mbox{\textit{\fontsize{11.2}{5}\selectfont errbench}}}
\newcommand{\Kassandra}{Kassandra}
\newcommand{\microprogram}{microprogram}
\newcommand{\microprograms}{microprograms}
\newcommand{\microinstruction}{microinstruction}
\newcommand{\microinstructions}{microinstructions}
\newcommand{\Microinstructions}{Microinstructions}
\newcommand{\writeSC}{\smalltt{write()}}
\newcommand{\fsyncSC}{\smalltt{fsync()}}
\newcommand{\msyncSC}{\smalltt{msync()}}
\newcommand{\fdatasyncSC}{\smalltt{x fdatasync()}}
\newcommand{\linkSC}{\smalltt{link()}}
\newcommand{\mkdirSC}{\smalltt{mkdir()}}
\newcommand{\fempty}{$\phi$}
\newcommand{\fexists}{$\surd$}
\newcommand{\creatSC}{{\smalltt{creat()}}}
\newcommand{\unlinkSC}{{\smalltt{unlink()}}}
\newcommand{\renameSC}{{\smalltt{rename()}}}
\newcommand\floor[1]{\lfloor#1\rfloor}
\newcommand\ceil[1]{\lceil#1\rceil}
\newcommand{\totbugs}{60}
\newcommand{\totapps}{11}
\newcommand{\totappsw}{eleven}
\newcommand*{\combination}[2]{{}^{#1}C_{#2}}

\newcommand{\used}{$\surd$}
\newcommand{\usedpar}{$P$}
\newcommand{\useddollar}{$\surd$\textsuperscript{\$}}
\newcommand{\usedadler}{$\surd$\textsuperscript{a}}
\newcommand{\usedparstar}{$P$\textsuperscript{$*$}}
\newcommand{\notused}{}
\newcommand{\numapps}{eight}

\if 0 % colored version
\newcommand{\yes}{$\surd$}
\newcommand{\yesi}{\colorbox{gray!30}{$\surd$\textsubscript{$i$}}}
\newcommand{\nolow}{\colorbox{gray!30}{$\times$\textsubscript{$l$}}}
\newcommand{\no}{\colorbox{gray!85}{$\times$}}
\newcommand{\complower}{$L$}
\newcommand{\compmoder}{$M$}
\newcommand{\comphigher}{\colorbox{gray!85}{$H$}}
\newcommand{\na}{\footnotesize na}
\fi

\newcommand{\yes}{$\surd$}
\newcommand{\yesi}{$\surd$\textsubscript{$i$}}
\newcommand{\nolow}{$\times$\textsubscript{$l$}}
\newcommand{\nomod}{$\times$\textsubscript{$m$}}
\newcommand{\no}{$\times$}
\newcommand{\yeslow}{$\surd$\textsuperscript{$l$}}
\newcommand{\complower}{$L$}
\newcommand{\compmoder}{$M$}
\newcommand{\comphigher}{$H$}
\newcommand{\na}{\footnotesize na}

\newcommand*{\termindex}[2]{$\langle$\textit{epoch}:{#1}, \textit{index}:{#2}$\rangle$}
\newcommand*{\epochindex}[2]{{#1}.{#2}}
\newcommand*{\termindexnovar}{$\langle$\textit{epoch}, \textit{index}$\rangle$}
\newcommand*{\snapid}{$\langle$\textit{snap-index}, \textit{chunk}\#$\rangle$}
\newcommand*{\rafttermindexnovar}{$\langle$\textit{term}, \textit{index}$\rangle$}
\newcommand{\quotes}[1]{``#1''}
\newcommand{\camera}[1]{\textcolor{Black}{#1}}
\newcommand{\addcamera}[1]{\textcolor{Black}{#1}}
\newcommand{\todo}[1]{\textcolor{Red}{#1}}

%% file: title.tex
\title{\vspace{0.2in}\Large \bf From WiscKey to Bourbon: A Learned Index for Log-Structured Merge Trees}

\author{
{\rm Yifan Dai, Yien Xu, Aishwarya Ganesan, Ramnatthan Alagappan, Brian Kroth$^\dagger$,}\\
{\rm Andrea C. Arpaci-Dusseau, and Remzi H. Arpaci-Dusseau}\\ \\[-1.0ex]
University of Wisconsin -- Madison ~~~$^\dagger$ Microsoft Gray Systems Lab
}
\date{}
\maketitle

%% file: abstract.tex
%\noindent\textbf{Abstract.1.5} abstract -- 1.8$\times$}
%Learned indexes use machine-learning models to index 

\noindent
\textbf{\textit{Abstract.}} We introduce \sysname, a log-structured merge (LSM) tree that utilizes machine learning to provide fast lookups. We base the design and implementation of \sysname\ on empirically-grounded principles that we derive through careful analysis of LSM design. \sysname\ employs greedy piecewise linear regression to learn key distributions, enabling fast lookup with minimal computation, and applies a cost-benefit strategy to decide when learning will be worthwhile. Through a series of experiments on both synthetic and real-world datasets, we show that \sysname\ improves lookup performance by 1.23$\times$-1.78$\times$ as compared to state-of-the-art production LSMs.

%% file: intro.tex
\section{Introduction} 
\label{sec-intro} 

% ML is amazing
Machine learning is transforming how we build computer applications and
systems. Instead of writing code in the traditional algorithmic mindset, one
can instead collect the proper data, train a model, and thus implement a
robust and general solution to the task at hand. This data-driven, empirical
approach has been called ``Software 2.0''~\cite{Karpathy17-Software2}, hinting
at a world where an increasing amount of the code we deploy is realized in
this manner; a number of landmark successes over the past decade lend credence
to this argument, in areas such as image~\cite{Krizhevsky12-DeepImgnet} and
speech recognition~\cite{Graves+13-DeepSpeech}, machine
translation~\cite{Wu+16-Translate}, game playing~\cite{Silver+16-Go}, and many
other areas~\cite{bojarski2016end,esteva2019guide,erfani2016high}.   

% Learned indexes specifically
% One place: replacing lookup structures.
One promising line of work, for using ML to
improve core systems is that of the ``learned
index''~\cite{learnedindex}. This approach applies machine learning to
supplant the traditional index structure found in database systems,
namely the ubiquitous B-Tree~\cite{Comer79-Btree}. To look up a key, the
system uses a learned function that predicts the location of the key (and
value); when successful, this approach can improve lookup
performance, in some cases significantly, and also potentially reduce space
overhead. Since this pioneering work, numerous follow ups~\cite{sagedb,fittree,ding2019alex}
have been proposed that use better models, better
tree structures, and generally improve how learning can
reduce tree-based access times and overheads.

% LSMs: important class, overlooked
However, one critical approach has not yet been transformed in this
``learned'' manner: the Log-structured Merge Tree
(LSM)~\cite{o1996log,sears2012blsm,wisckey}. LSMs were introduced in the late
\camera{'90s}, gained popularity a decade later through work at
Google on BigTable~\cite{Chang+06-BigTable} and
LevelDB~\cite{Ghemawat+11-LevelDB}, and have become widely used in industry,
including in Cassandra~\cite{Lakshman+09-Cassandra},
RocksDB~\cite{rocksdb}, and many other
systems~\cite{riak,george2011hbase}. LSMs have many positive properties as
compared to B-trees and their cousins, including high insert performance~\cite{pebblesdb,wisckey,dayan2017monkey}.

% What we do
In this paper, we apply the idea of the learned index to LSMs. A major challenge is that while learned indexes are primarily tailored for read-only settings, LSMs are optimized for writes. Writes cause disruption to learned indexes because models learned over existing data must now be updated to accommodate the changes; the system thus must re-learn the data repeatedly. However, we find that LSMs are well-suited for learned indexes. For example, although writes modify the LSM, most portions of the tree are immutable; thus, learning a function to predict key/value locations can be done once, and used as long as the immutable data lives. However, many challenges arise. For example, variable key or value sizes make learning a function to predict locations more difficult, and performing model building too soon may lead to significant resource waste.

Thus, we first study how an existing LSM system, WiscKey~\cite{wisckey}, functions in great detail (\sref{sec-good-match}). We focus on WiscKey because it is a state-of-the-art LSM implementation that is significantly faster than LevelDB and RocksDB~\cite{wisckey}. Our analysis leads to many interesting insights from which we develop five {\em learning guidelines}: a set of rules that aid an LSM system to successfully incorporate learned indexes. For example, while it is useful to learn the stable, low levels in an LSM, learning higher levels can yield benefits as well because lookups must always search the higher levels. Next, not all files are equal: some files even in the lower levels are very short-lived; a system must avoid learning such files, or resources can be wasted. Finally, workload- and data-awareness is important; based on the workload and how the data is loaded, it may be more beneficial to learn some portions of the tree than others.  

We apply these learning guidelines to build \sysname, a learned-index implementation of WiscKey (\sref{sec-llsm-design}). \sysname\ uses piece-wise linear regression, a simple but effective model that enables both fast training (i.e., learning) and inference (i.e., lookups) with little space overhead. \sysname\ employs {\em file learning}: models are built over files given that an LSM file, once created, is never modified in-place. \sysname\ implements a cost-benefit analyzer that dynamically decides whether or not to learn a file, reducing unnecessary learning while maximizing benefits. While most of the prior work on learned indexes~\cite{learnedindex, fittree, ding2019alex} has made strides in optimizing stand-alone data structures, \sysname\ integrates learning into a production-quality system that is already highly optimized. \camera{\sysname's implementation adds around 5K LOC to WiscKey (which has $\sim$20K LOC).}

We analyze the performance of \sysname\ on a range of synthetic and real-world datasets and workloads (\sref{sec-eval}). We find that \sysname\ reduces the indexing costs of WiscKey significantly and thus offers 1.23$\times$ -- 1.78$\times$ faster lookups for various datasets. Even under workloads with significant write load, \sysname\ speeds up a large fraction of lookups and, through cost-benefit, avoids unnecessary (early) model building. Thus, \sysname\ matches the performance of an aggressive-learning approach but performs model building more judiciously. Finally, most of our analysis focuses on the case where fast lookups will make the most difference, namely when the data resides in memory (i.e., in the file-system page cache). However, we also experiment with \sysname\ when data resides on a fast storage device (an Optane SSD) or when data does not fit entirely in memory, and show that benefits can still be realized.

This paper makes four contributions. We present the first detailed study of how LSMs function internally with learning in mind. We formulate a set of guidelines on how to integrate learned indexes into an LSM (\sref{sec-good-match}). We present the design and implementation of \sysname\ which incorporates learned indexes into a real, highly optimized, production-quality LSM system (\sref{sec-llsm-design}). Finally, we analyze \sysname's performance in detail, and demonstrate its benefits (\sref{sec-eval}).

%% file: bg.tex
\section{Background}
\label{sec-overview}

We first describe log-structured merge trees and explain how data is organized in LevelDB. Next, we describe WiscKey, a modified version of LevelDB that we adopt as our baseline. We then provide a brief background on learned indexes.%and analyze if they can be applied to LSMs to make lookups faster.

\subsection{LSM and LevelDB}

An LSM tree is a persistent data structure used in key-value stores to support efficient inserts and updates~\cite{o1996log}. Unlike B-trees that require many random writes to storage upon updates, LSM trees perform writes sequentially, thus achieving high write throughput~\cite{o1996log}. 

An LSM organizes data in multiple {\em levels}, with the size of each level increasing exponentially. Inserts are initially buffered in an in-memory structure; once full, this structure is merged with the first level of on-disk data. This procedure resembles merge-sort and is referred to as compaction. Data from an on-disk level is also merged with the successive level if the size of the level exceeds a limit. Note that updates do not modify existing records in-place; they follow the same path as inserts. As a result, many versions of the same item can be present in the tree at a time. Throughout this paper, we refer to the levels that contain the newer data as {\em higher} levels and the older data as {\em lower} levels.

A lookup request must return the latest version of an item. Because higher levels contain the newer versions, the search starts at the topmost level. First, the key is searched for in the in-memory structure; if not found, it is searched for in the on-disk tree starting from the highest level to the lowest one. The value is returned once the key is found at a level.  

\input{fig-leveldb}

LevelDB~\cite{Ghemawat+11-LevelDB} is a widely used key-value store built using LSM. Figure~\ref{fig-leveldb}(a) shows how data is organized in LevelDB. A new key-value pair is first written to the {\em memtable}; when full, the memtable is converted into an immutable table which is then compacted and written to disk sequentially as {\em sstables}. The sstables are organized in seven levels ($L_0$ being the highest level and $L_6$ the lowest) \camera{and each sstable corresponds to a file}. LevelDB ensures that key ranges of different sstables at a level are disjoint (two files will not contain overlapping ranges of keys); $L_0$ is an exception where the ranges can overlap across files. The amount of data at each level increases by a factor of ten; for example, the size of $L_1$ is 10MB, while $L_6$ contains several 100s of GBs. If a level exceeds its size limit, one or more sstables from that level are merged with the next level; this is repeated until all levels are within their limits.  %LevelDB stores the large volume of data at lower levels in many small files, not few very large files.

\noindent
\textbf{Lookup steps.} Figure~\ref{fig-leveldb}(a) also shows how a lookup request for key $k$ proceeds. \circled{1} {\em FindFiles}: If the key is not found in the in-memory tables, LevelDB finds the set of candidate sstable files that may contain $k$. A key in the worst case may be present in all $L_0$ files (because of overlapping ranges) and within one file at each successive level. \circled{2} {\em LoadIB+FB}: In each candidate sstable, an index block and a bloom-filter block are first loaded from the disk. \circled{3} {\em SearchIB}: The index block is binary searched to find the data block that may contain $k$. \circled{4} {\em SearchFB}: The filter is queried to check if $k$ is present in the data block. \circled{5} {\em LoadDB}: If the filter indicates presence, the data block is loaded. \circled{6} {\em SearchDB}: The data block is binary searched. \circled{7} {\em ReadValue}: If the key is found in the data block, the associated value is read and the lookup ends. If the filter indicates absence or if the key is not found in the data block, the search continues to the next candidate file. Note that blocks are not always loaded from the disk; index and filter blocks, and frequently accessed data blocks are likely to be present in memory (i.e., file-system cache).

We refer to these search steps at a level that occur as part of a single lookup as an {\em internal lookup}. A single lookup thus consists of many internal lookups. A {\em negative internal lookup} does not find the key, while a {\em positive internal lookup} finds the key and is thus the last step of a lookup request.  

We differentiate indexing steps from data-access steps; indexing steps such as {\em FindFiles}, {\em SearchIB}, {\em SearchFB}, and {\em SearchDB} search through the files and blocks to find the desired key, while data-access steps such as {\em LoadIB+FB}, {\em LoadDB}, and {\em ReadValue} read the data from storage. Our goal is to reduce the time spent in indexing.

\subsection{WiscKey}

In LevelDB, compaction results in large write amplification because {\em both} keys and values are sorted and rewritten. Thus, LevelDB suffers from high compaction overheads, affecting foreground workloads. 

WiscKey~\cite{wisckey} (and Badger~\cite{badgerdb}) reduces this overhead by storing the values separately; the sstables contain only keys and pointers to the values as shown in Figure~\ref{fig-leveldb}(b). With this design, compaction sorts and writes only the keys, leaving the values undisturbed, thus reducing I/O amplification and overheads. WiscKey thus performs significantly better than other optimized LSM implementations such as LevelDB and RocksDB. Given these benefits, we adopt WiscKey as the baseline for our design. Further, WiscKey's key-value separation enables our design to handle variable-size records; we describe how in more detail in \sref{ssec-llsm-design-wisckey}.    

The write path of WiscKey is similar to that of LevelDB except that values are written to a {\em value log}. A lookup in WiscKey also involves searching at many levels and a final read into the log once the target key is found. The size of WiscKey's LSM tree is much smaller than LevelDB because it does not contain the values; hence, it can be entirely cached in memory~\cite{wisckey}. Thus, a lookup request involves multiple searches in the in-memory tree, and the {\em ReadValue} step performs one final read to retrieve the value.

\subsection{Optimizing Lookups in LSMs}

Performing a lookup in LevelDB and WiscKey requires searching at multiple levels. Further, within each sstable, many blocks are searched to find the target key. Given that LSMs form the basis of many embedded key-value stores (e.g., LevelDB, RocksDB~\cite{rocksdb}) and distributed storage systems (e.g., BigTable~\cite{Chang+06-BigTable}, Riak~\cite{riak}), optimizing lookups in LSMs can have huge benefits.

A recent body of work, starting with learned indexes~\cite{learnedindex}, makes a case for replacing or augmenting traditional index structures with machine-learning models. The key idea is to train a model (such as linear regression or neural nets) on the input so that the model can predict the position of a record in the sorted dataset. The model can have inaccuracies, and thus the prediction has an associated error bound. During lookups, if the model-predicted position of the key is correct, the record is returned; if it is wrong, a local search is performed within the error bound. For example, if the predicted position is $pos$ and the minimum and maximum error bounds are $\delta$$_m$$_i$$_n$ and $\delta$$_m$$_a$$_x$, then upon a wrong prediction, a local search is performed between $pos-\delta$$_m$$_i$$_n$ and $pos+\delta$$_m$$_a$$_x$.

%our first contribution in this paper is to answer
Learned indexes can make lookups significantly faster. Intuitively, a learned index turns a $O(log$-$n)$ lookup of a B-tree into a $O$$(1)$ operation. Empirically, learned indexes provide 1.5$\times$ -- 3$\times$ faster lookups than B-trees~\cite{learnedindex}. Given these benefits, we ask the following questions: \camera{{\em can learned indexes for LSMs make lookups faster? If yes, under what scenarios?}} 

Traditional learned indexes do not support updates because models learned over the existing data would change with modifications~\cite{learnedindex, ding2019alex, fittree}. However, LSMs are attractive for their high performance in write-intensive workloads because they perform writes only sequentially. Thus, we examine: \camera{{\em how to realize the benefits of learned indexes while supporting writes for which LSMs are optimized?}} We answer these two questions next.

%% file: fig-leveldb.tex
\begin{figure}[t!]
\centering
\includegraphics[scale=0.68]{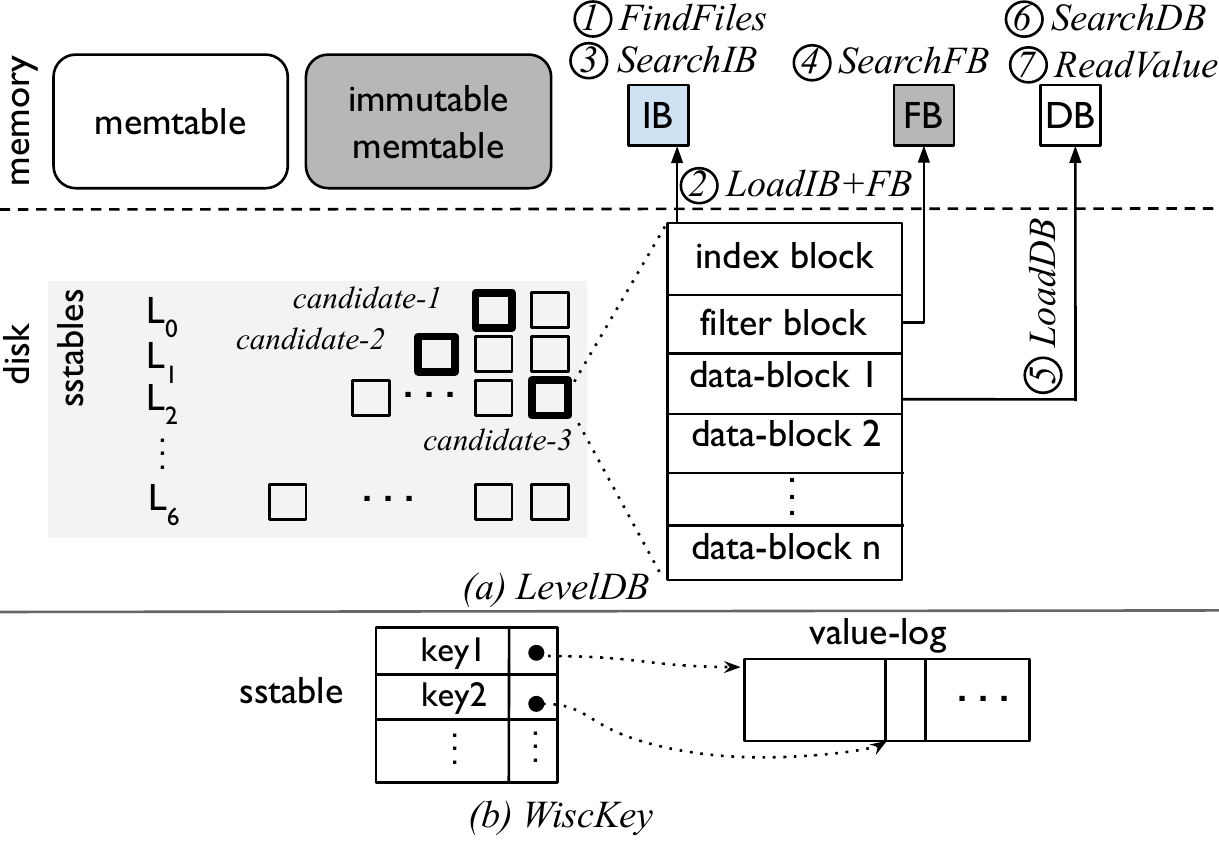}
\vspace{-0.21in}
\mycaption{fig-leveldb}{LevelDB and WiscKey}{(a) shows how data is organized in LevelDB and how a lookup is processed. The search in in-memory tables is not shown. The candidate sstables are shown in bold boxes. (b) shows how keys and values are separated in WiscKey.}
\end{figure}

%% file: mot.tex
\section{Learned Indexes: a Good Match for LSMs?}
\label{sec-good-match}

In this section, we first analyze if learned indexes could be beneficial for LSMs and examine under what scenarios they can improve lookup performance. We then provide our intuition as to why learned indexes might be appropriate for LSMs even when allowing writes. We conduct an in-depth study based on measurements of how WiscKey functions internally under different workloads to validate our intuition. From our analysis, we derive a set of learning guidelines.

\subsection{Learned Indexes: Beneficial Regimes}

\input{fig-seq-regime}

A lookup in LSM involves several indexing and data-access steps. Optimized indexes such as learned indexes can reduce the overheads of indexing but cannot reduce data-access costs. In WiscKey, learned indexes can thus potentially reduce the costs of indexing steps such as {\em FindFiles}, {\em SearchIB}, and {\em SearchDB}, while data-access costs (e.g.,  {\em ReadValue}) cannot be significantly reduced. As a result, learned indexes can improve overall lookup performance if indexing contributes to a sizable portion of the total lookup latency. We identify scenarios where this is the case.   

First, when the dataset or a portion of it is cached in memory, data-access costs are low, and so indexing costs become significant. Figure~\ref{fig-regime} shows the breakdown of lookup latencies in WiscKey. The first bar shows the case when the dataset is cached in memory; the second bar shows the case where the data is stored on a flash-based SATA SSD. With caching, data-access and indexing costs contribute almost equally to the latency. Thus, optimizing the indexing portion can reduce lookup latencies by about 2$\times$. When the dataset is not cached, data-access costs dominate and thus optimizing indexes may yield smaller benefits (about 20\%).

However, learned indexes are not limited to scenarios where data is cached in memory. They offer benefit on fast storage devices that are currently prevalent and can do more so on emerging faster devices. The last three bars in Figure~\ref{fig-regime} show the breakdown for three kinds of devices: flash-based SSDs over SATA and NVMe, and an Optane SSD. As the device gets faster, lookup latency (as shown at the top) decreases, but the fraction of time spent on indexing increases. For example, with SATA SSDs, indexing takes about 17\% of the total time; in contrast, with Optane SSDs, indexing takes 44\% and thus optimizing it with learned indexes can potentially improve performance by 1.8$\times$. More importantly, the trend in storage performance favors the use of learned indexes. With storage performance increasing rapidly and emerging technologies like 3D Xpoint memory providing very low access latencies, indexing costs will dominate and thus learned indexes will yield increasing benefits.

\noindent
\textbf{Summary.} Learned indexes could be beneficial when the database or a portion of it is cached in memory. With fast storage devices, regardless of caching, indexing contributes to a significant fraction of the lookup time; thus, learned indexes can prove useful in such cases. With storage devices getting faster, learned indexes will be even more beneficial.

\subsection{Learned Indexes with Writes}

\input{fig-lifetime-distributions} 
\input{fig-lookups-distribution}

Learned indexes provide higher lookup performance compared to traditional indexes for read-only analytical workloads. However, a major drawback of learned indexes (as described in ~\cite{learnedindex}) is that they do not support modifications such as inserts and updates~\cite{fittree, ding2019alex}. The main problem with modifications is that they alter the data distribution and so the models must be re-learned; for write-heavy workloads, models must be rebuilt often, incurring high overheads.

At first, it may seem like learned indexes are not a good match for write-heavy situations for which LSMs are optimized. However, we observe that the design of LSMs fits well with learned indexes. Our key realization is that although updates can change portions of the LSM tree, a large part remains immutable. Specifically, newly modified items are buffered in the in-memory structures or present in the higher levels of the tree, while stable data resides at the lower levels. Given that a large fraction of the dataset resides in the stable, lower levels, lookups to this fraction can be made faster with no or few re-learnings. In contrast, learning in higher levels may be less beneficial: they change at a faster rate and thus must be re-learned often.

We also realize that the immutable nature of sstable files makes them an ideal unit for learning. Once learned, these files are never updated and thus a model can be useful until the file is replaced. Further, the data within an sstable is sorted; such sorted data can be learned using simple models. A level, which is a collection of many immutable files, can also be learned as a whole using simple models. The data in a level is also sorted: the individual sstables are sorted, and there are no overlapping key ranges across sstables.

We next conduct a series of in-depth measurements to validate our intuitions. Our experiments 
confirm that while a part of our intuition is indeed true, there are some subtleties (for example, in learning files at higher levels). Based on these experimental results, we formulate a set of {\em learning guidelines}: a few simple rules that an LSM that applies learned indexes should follow.

\noindent
\textbf{Experiments: goal and setup.} The goal of our experiments is to determine how long a model will be useful and how often it will be useful. A model built for a sstable file is useful as long as the file exists; thus, we first measure and analyze sstable lifetimes. How often a model will be used is determined by how many internal lookups it serves; thus, we next measure the number of internal lookups to each file. Since models can also be built for entire levels, we finally measure level lifetimes as well. To perform our analysis, we run workloads with varying amounts of writes and reads, and measure the lifetimes and number of lookups. We conduct our experiments on WiscKey, but we believe our results are applicable to most LSM implementations. We first load the database with 256M key-value pairs. We then run a workload with a single rate-limited client that performs 200M operations, a fraction of which are writes. Our workload chooses keys uniformly at random.%: every key is equally likely to be accessed. 

\renewcommand{\thefootnote}{\fnsymbol{footnote}}

%\vspace{0.2in}
\noindent
\textbf{Lifetime of SSTables.} To determine how long a model will be useful, we first measure and analyze the lifetimes of sstables. To do so, we track the creation and deletion times of all sstables. For files created during the load phase, we assign the workload-start time as their creation time; for other files, we record the actual creation times. If the file is deleted during the workload, then we calculate its exact lifetime. However, some files are not deleted by the end of the workload and we must estimate their lifetimes.\footnote[2]{If the files are created during load, we assign the workload duration as their lifetimes. If not, we estimate the lifetime of a file based on its creation time ($c$) and the total workload time ($w$); the lifetime of the file is at least $w-c$. We thus consider the lifetime distribution of other files that have a lifetime of at least $w-c$. We then pick a random lifetime in this distribution and assign it as this file's lifetime.}

Figure~\ref{fig-lifetime-distributions}(a) shows the average lifetime of sstable files at different levels. We make three main observations. First, the average lifetime of sstable files at lower levels is greater than that of higher levels. Second, at lower percentages of writes, even files at higher levels have a considerable lifetime; for example, at 5\% writes, files at $L_0$ live for about 2 minutes on an average. Files at lower levels live much longer; files at $L_4$ live about 150 minutes. Third, although the average lifetime of files reduces with more writes, even with a high amount of writes, files at lower levels live for a long period. For instance, with 50\% writes, files at $L_4$ live for about 60 minutes. In contrast, files at higher level live only for a few seconds; for example, an $L_0$ file lives only about 10 seconds.

We now take a closer look at the lifetime distribution. Figure~\ref{fig-lifetime-distributions}(b) shows the distributions for $L_1$ and $L_4$ files with 5\% writes. We first note that some files are very short-lived, while some are long-lived. For example, in $L_1$, the lifetime of about 50\% of the files is only about 2.5 seconds. If files cross this threshold, they tend to live for much longer times; almost all of the remaining $L_1$ files live over five minutes.

Surprisingly, even at $L_4$, which has a higher average lifetime for files, a few files are very short-lived. We observe that about 2\% of $L_4$ files live less than a second. We find that there are two reasons why a few files live for a very short time. First, compaction of a $L_i$ file creates a new file in $L_i$$_+$$_1$ which is again immediately chosen for compaction to the next level. Second, compaction of a $L_i$ file creates a new file in $L_i$$_+$$_1$, which has overlapping key ranges with the next file that is being compacted from $L_i$. Figure~\ref{fig-lifetime-distributions}(c) shows that this pattern holds for other percentages of writes too. We observed that this holds for other levels as well. From the above observations, we arrive at our first two learning guidelines.

\noindent
\textbf{\textit{Learning guideline - 1: Favor learning files at lower levels.}} Files at lower levels live for a long period even for high write percentages; thus, models for these files can be used for a long time and need not be rebuilt often.

\noindent
\textbf{\textit{Learning guideline - 2: Wait before learning a file.}} A few files are very short-lived, even at lower levels. Thus, learning must be invoked only after a file has lived up to a threshold lifetime after which it is highly likely to live for a long time. 

%\vspace{0.2in}
\noindent
\textbf{Internal Lookups at Different Levels.} To determine how many times a model will be used, we analyze the number of lookups served by the sstable files. We run a workload and measure the number of lookups served by files at each level and plot the average number of lookups per file at each level. Figure~\ref{fig-lookups-distribution}(a) shows the result when the dataset is loaded in an uniform random order. The number of internal lookups is higher for higher levels, although a large fraction of data resides at lower levels. This is because, at higher levels, many internal lookups are negative, as shown in Figure~\ref{fig-lookups-distribution}(a)(ii). The number of positive internal lookups is as expected: higher in lower levels as shown in Figure~\ref{fig-lookups-distribution}(a)(iii). This result shows that files at higher levels serve many negative lookups and thus are worth optimizing. While bloom filters may already make these negative lookups faster, the index block still needs to be searched (before the filter query).

\camera{We also conduct the same experiment with another workload where the access pattern follows a zipfian distribution (most requests are to a small set of keys). Most of the results exhibit the same trend as the random workload except for the number of positive internal lookups, as shown in Figure~\ref{fig-lookups-distribution}(a)(iv). Under the zipfian workload, higher level files also serve numerous positive lookups, because the workload accesses a small set of keys which are often updated and thus stored in higher levels.}

Figure~\ref{fig-lookups-distribution}(b) shows the result when the dataset is sequentially loaded, i.e., keys are inserted in ascending order. In contrast to the randomly-loaded case, there are no negative lookups because keys of different sstable files do not overlap even across levels; the {\em FindFiles} step finds the one file that may contain the key. Thus, lower levels serve more lookups and can have more benefits from learning. From these observations, we arrive at the next two learning guidelines.    

\noindent
\textbf{\textit{Learning guideline - 3: Do not neglect files at higher levels.}} Although files at lower levels live longer and serve many lookups, files at higher levels can still serve many negative lookups \camera{and in some cases, even many positive lookups}. Thus, learning files at higher levels can make \camera{both} internal lookups faster.

\noindent
\textbf{\textit{Learning guideline - 4: Be workload- and data-aware.}} Although most data resides in lower levels, if the workload does not lookup that data, learning those levels will yield less benefit; learning thus must be aware of the workload. Further, the order in which the data is loaded influences which levels receive a large fraction of internal lookups; thus, the system must also be data-aware. The amount of internal lookups acts as a proxy for both the workload and load order. Based on the amount of internal lookups, the system must dynamically decide whether to learn a file or not.

%\vspace{0.2in}
\noindent
\textbf{Lifetime of Levels.} Given that a level as a whole can also be learned, we now measure and analyze the lifetimes of levels. Level learning cannot be applied at $L_0$ because it is unsorted: files in $L_0$ can have overlapping key ranges. Once a level is learned, any change to the level causes a re-learning. A level changes when new sstables are created at that level, or existing ones are deleted. Thus, intuitively, a level would live for an equal or shorter duration than the individual sstables. However, learning at the granularity of a level has the benefit that the candidate sstables need not be found in a separate step; instead, upon a lookup, the model just outputs the sstable and the offset within it. 

We examine the changes to a level by plotting the timeline of file creations and deletions at $L_1$, $L_2$, $L_3$, and $L_4$ in Figure~\ref{fig-level-tl}(a) for a 5\%-write workload; we do not show $L_0$ for the reason above. On the y-axis, we plot the number of changes divided by the total files present at that level. A value of 0 means there are no changes to the level; a model learned for the level can be used as long as the value remains 0. A value greater than $0$ means that there are changes in the level and thus the model has to re-learned. Higher values denote a larger fraction of files are changed.

\input{fig-level-tl}

First, as expected, we observe that the fraction of files that change reduces as we go down the levels because lower levels hold a large volume of data in many files, confirming our intuition. We also observe that changes to levels arrive in bursts. These bursts are caused by compactions that cause many files at a level to be rewritten. Further, these bursts occur at almost the same time across different levels. The reason behind this is that for the dataset we use, levels $L_0$ through $L_3$ are full and thus any compaction at one layer results in cascading compactions which finally settle at the non-full $L_4$ level. The levels remain static between these bursts. The duration for which the levels remain static is longer with a lower amount of writes; for example, with 5\% writes, as shown in the figure, this period is about 5 minutes. However, as the amount of writes increases, the lifetime of a level reduces as shown in Figure~\ref{fig-level-tl}(b); for instance, with 50\% writes, the lifetime of $L_4$ reduces to about 25 seconds. From these observations, we arrive at our final learning guideline.

\noindent
\textbf{\textit{Learning guideline - 5: Do not learn levels for write-heavy workloads.}} Learning a level as a whole might be more appropriate when the amount of writes is very low or if the workload is read-only. For write-heavy workloads, level lifetimes are very short and thus will induce frequent re-learnings.

\noindent
\textbf{Summary.} We analyzed how LSMs behave internally by measuring and analyzing the lifetimes of sstable files and levels, and the amount of lookups served by files at different levels. From our analysis, we derived five learning guidelines. We next describe how we incorporate the learning guidelines in an LSM-based storage system. 

%% file: fig-seq-regime.tex
\begin{figure}[t!]
\centering
\includegraphics[scale=1.1]{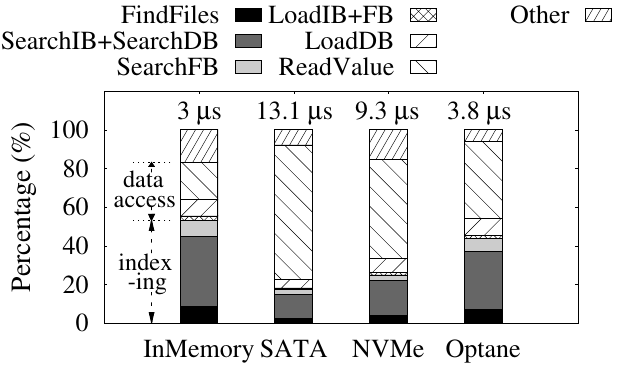}
\vspace{-0.08in}
\mycaption{fig-regime}{Lookup Latency Breakdown}{The figure shows the breakdown of lookup latency in WiscKey. The first bar shows the case when data is cached in memory. The other three bars show the case where the dataset is stored on different types of SSDs. We perform 10M random lookups on the Amazon Reviews dataset~\cite{ardataset}; the figure shows the breakdown of the average latency (shown at the top of each bar). The indexing portions are shown in solid colors; data access and other portions are shown in patterns.}
\end{figure}

%the figure shows the breakdown of the average latency. The indexing portions are shown in solid colors; data access and other portions are shown in patterns. Numbers on the top show the average latency of a lookup.

%% file: fig-lifetime-distributions.tex
\newcommand{\rulesep}{\unskip\ \vrule\ }
\begin{figure*}[t!] 

\begin{center}
\includegraphics[scale=0.83]{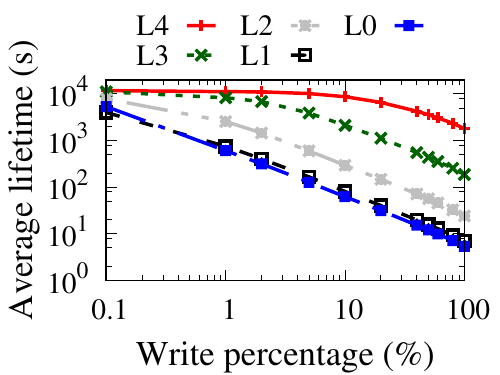} \hspace{0.06in}
\vrule \hspace{0.06in}
\includegraphics[scale=0.83]{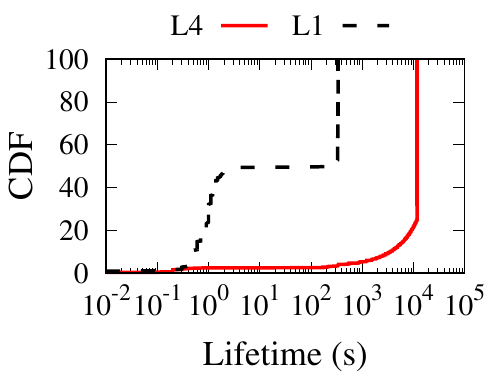} \hspace{0.06in}
\vrule  \hspace{0.06in}
\includegraphics[scale=0.76]{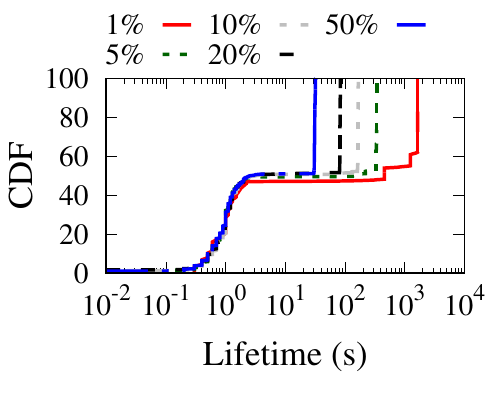}
\includegraphics[scale=0.76]{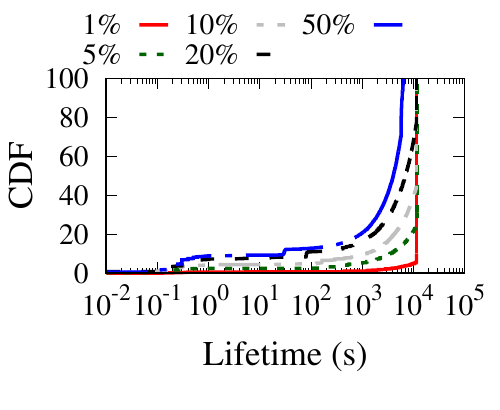}
\end{center}
\vspace{-0.3in}
\hspace{4.45in}{\footnotesize (i) Level 1}\hspace{1.1in}{\footnotesize (ii) Level 4}

\hspace{0.01in}{\footnotesize (a) Average lifetimes with varying write \%}\hspace{0.03in}{\footnotesize(b) Lifetime distribution with 5\% writes}\hspace{0.65in}{\footnotesize (c) Lifetime distributions with varying write \%}
%\hspace{-0.1in}{\footnotesize(a)(i) Synchronous Replication - Redis}\hspace{0.45in}{\footnotesize(a)(ii) Synchronous Replication - ZooKeeper}\hspace{0.45in}{\footnotesize(b) Asynchronous Replication - Redis}
%\hspace{0.5in}{\footnotesize(a) ZooKeeper}\hspace{1.3in}{\footnotesize(b) Redis (Synchronous Replication)}\hspace{0.45in}{\footnotesize(c) Redis (Asynchronous Replication)}
%\vspace{-0.03in}
\mycaption{fig-lifetime-distributions}{SSTable Lifetimes}{(a) shows the average lifetime of sstable files in levels $L_4$ to $L_0$. (b) shows the distribution of lifetimes of sstables in $L_1$ and $L_4$ with 5\% writes. (c) shows the distribution of lifetimes of sstables for different write percentages in $L_1$ and $L_4$.}

\end{figure*}

%% file: fig-lookups-distribution.tex
\begin{figure*}[t!] 
\begin{center}
\includegraphics[scale=0.68]{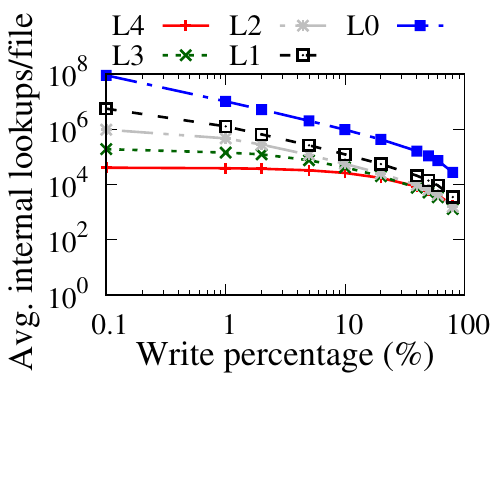}
\includegraphics[scale=0.68]{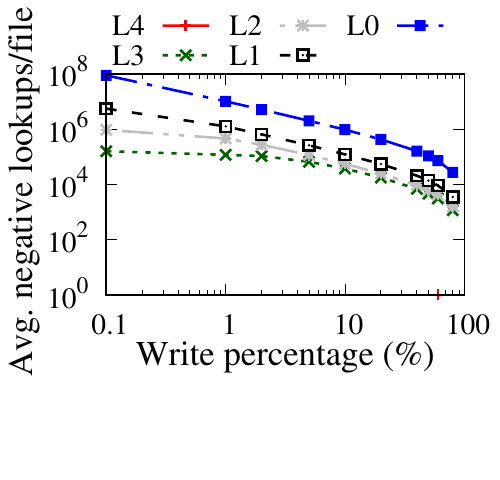}
\includegraphics[scale=0.68]{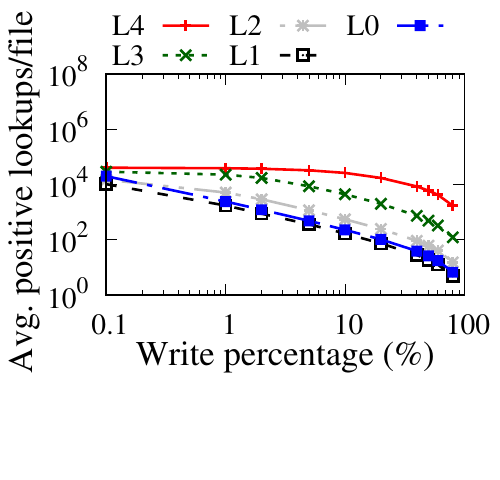}
\includegraphics[scale=0.68]{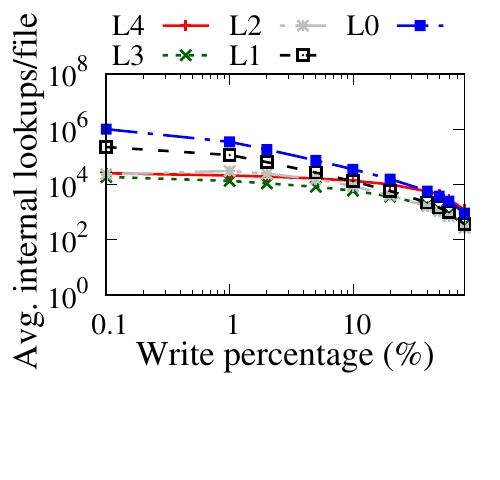}
\rulesep
\includegraphics[scale=0.68]{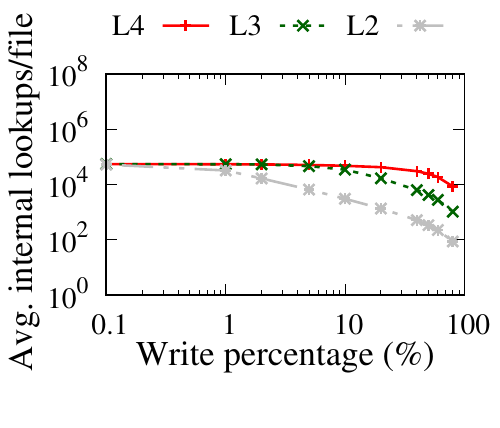}
\end{center}
\vspace{-0.43in}
\hspace{0.55in}{\footnotesize (i) Total}\hspace{0.95in}{\footnotesize(ii) Negative}\hspace{0.9in}{\footnotesize(iii) Positive}\hspace{0.65in}{\footnotesize(iv) Positive (Zipfian)}

\hspace{2in}{\footnotesize (a) Randomly loaded dataset}\hspace{2.35in}{\footnotesize(b) Sequentially loaded dataset}
%\hspace{-0.1in}{\footnotesize(a)(i) Synchronous Replication - Redis}\hspace{0.45in}{\footnotesize(a)(ii) Synchronous Replication - ZooKeeper}\hspace{0.45in}{\footnotesize(b) Asynchronous Replication - Redis}
%\hspace{0.5in}{\footnotesize(a) ZooKeeper}\hspace{1.3in}{\footnotesize(b) Redis (Synchronous Replication)}\hspace{0.45in}{\footnotesize(c) Redis (Asynchronous Replication)}
\vspace{-0.15in}
\mycaption{fig-lookups-distribution}{Number of Internal Lookups Per File}{(a)(i) shows the average internal lookups per file at each level for a randomly loaded dataset. (b) shows the same for sequentially loaded dataset. (a)(ii) and (a)(iii) show the negative and positive internal lookups for the randomly loaded case. \camera{(a)(iv) shows the positive internal lookups for the randomly loaded case when the workload distribution is Zipfian.}}
\end{figure*}

%% file: fig-level-tl.tex
\begin{figure}[t!]
\centering
\captionsetup[subfigure]{justification=centering}
\subfloat[Timeline of changes]
{
\includegraphics[scale=0.24]{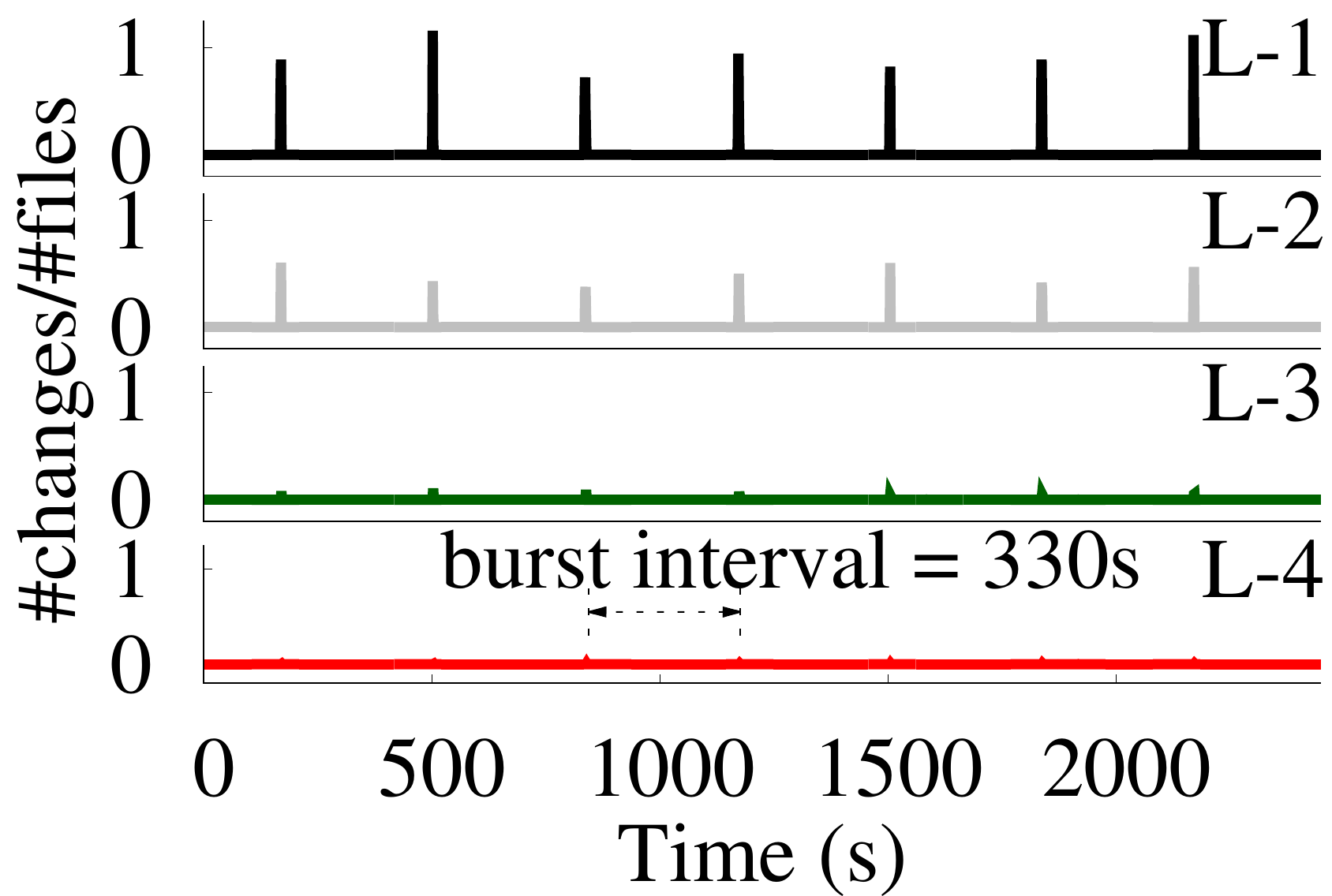}
}
\subfloat[Time between bursts for L4]
{
\includegraphics[scale=0.63]{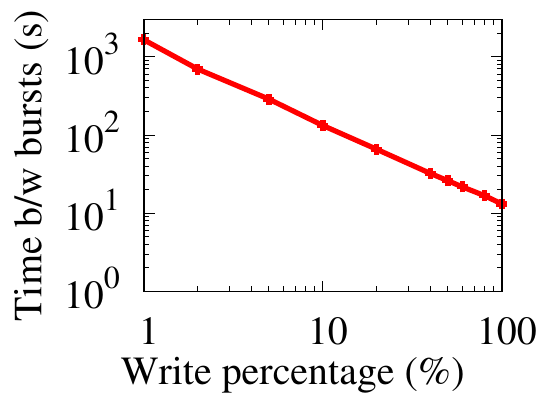}
}
\vspace{-0.05in}
\mycaption{fig-level-tl}{Changes at Levels}{(a) shows the timeline of file creations and deletions at different levels. Note that \#changes/\#files is higher than 1 in $L_1$ as there are more creations and deletions than the number of files. (b) shows the time between bursts for L4 for different write percentages.}
\end{figure}

%the scale for $L_1$ is different to accommodate this. The numbers in blue show the average number of files that do not change between two bursts.

%% file: llsm.tex
\section{Bourbon Design}
\label{sec-llsm-design}

We now describe \sysname, an LSM-based store that uses learning to make indexing faster. We first describe the model that \sysname\ uses to learn the data (\sref{ssec-model}). Then, we discuss how \sysname\ supports variable-size values (\sref{ssec-llsm-design-wisckey}) and its basic learning strategy (\sref{ssec-level-or-file}). We finally explain \sysname's cost-benefit analyzer that dynamically makes learning decisions to maximize benefit while reducing cost (\sref{ssec-cba}).   

\subsection{Learning the Data}
\label{ssec-model}

As we discussed, data can be learned at two granularities: individual sstables or levels. Both these entities are sorted datasets. The goal of a model that tries to learn the data is to predict the location of a key in such a sorted dataset. For example, if the model is constructed for a sstable file, it would predict the file offset given a key. Similarly, a level model would output the target sstable file and the offset within it.

Our requirements for a model is that it must have low overheads during learning and during lookups. Further, we would like the space overheads of the model to be small. We find that piecewise linear regression (PLR)~\cite{acharya2016fast,keogh2001online} satisfies these requirements well; thus, \sysname\ uses PLR to model the data. The intuition behind PLR is to represent a sorted dataset with a number of line segments. PLR constructs a model with an error bound; that is, each data point $d$ is guaranteed to lie within the range [$d_{pos}$ $-$ $\delta$, $d_{pos}$ $+$ $\delta$], where $d_{pos}$ is the predicted position of $d$ in the dataset and $\delta$ is the error bound specified beforehand. %The number of line segments is determined in a way such that all data points lie the error bound.   

To train the PLR model, \sysname\ uses the Greedy-PLR algorithm~\cite{xie2014maximum}. Greedy-PLR processes the data points one at a time; if a data point cannot be added to the current line segment without violating the error bound, then a new line segment is created and the data point is added to it. At the end, Greedy-PLR produces a set of line segments that represents the data. Greedy-PLR runs in linear time with respect to the number of data points. 

Once the model is learned, inference is quick: first, the correct line segment that contains the key is found (using binary search); within that line segment, the position of the target key is obtained by multiplying the key with the line's slope and adding the intercept. If the key is not present in the predicted position, a local search is done in the range determined by the error bound. Thus, lookups take $O(log$-$s)$ time, where $s$ is the number of segments, in addition to a constant time to do the local search. The space overheads of PLR are small: a few tens of bytes for every line segment.

\camera{Other models or algorithms such as RMI~\cite{learnedindex}, PGM-Index~\cite{Ferragina_2020}, or splines~\cite{kipf2020radixspline} may also be suitable for LSMs and may offer more benefits than PLR. We leave their exploration within LSMs for future work.}

\subsection{Supporting Variable-size Values}
\label{ssec-llsm-design-wisckey}

Learning a model that predicts the offset of a key-value pair is much easier if the key-value pairs are the same size. The model then can multiply the predicted position of a key by the size of the pair to produce the final offset. However, many systems allow keys and values to be of arbitrary sizes. 

\sysname\ requires keys to be of a fixed size, while values can be of any size. We believe this is a reasonable design choice because most datasets have fixed-size keys (e.g., user-ids are usually 16 bytes), while value sizes vary significantly. Even if keys vary in size, they can be padded to make all keys of the same size. \sysname\ supports variable-size values by borrowing the idea of key-value separation from WiscKey~\cite{wisckey}. With key-value separation, sstables in \sysname\ just contain the keys and the pointer to the values; values are maintained in the value log separately. With this, \sysname\ obtains the offset of a required key-value pair by getting the predicted position from the model and multiplying it with the record size (which is $keysize$ + $pointersize$.) The value pointer serves as the offset into the value log from which the value is finally read.   

\input{tbl-granularity}

\subsection{Level vs. File Learning}
\label{ssec-level-or-file}

\sysname\ can learn individual sstables files or entire levels. Our analysis in the previous section showed that files live longer than levels under write-heavy workloads, hinting that learning at the file granularity might be the best choice. We now closely examine this tradeoff to design \sysname's basic learning strategy. To do so, we compare the performance of file learning and level learning for different workloads. We initially load a dataset and build the models. For the read-only workload, the models need not be re-learned. In the mixed workloads, the models are re-learned as data changes. The results are shown in Table~\ref{tbl-granularity}.  

For mixed workloads, level learning performs worse than file learning. For a write-heavy (50\%-write) workload, with level learning, only a small percentage of internal lookups are able to use the model because with a steady stream of incoming writes, the system is unable to learn the levels. Only a mere 1.5\% of internal lookups take the model path; these lookups are the ones performed just after loading the data and when the initial level models are available. We observe that all the 66 attempted level learnings failed because the level changed before the learning completed. Because of the additional cost of re-learnings, level learning performs even worse than the baseline with 50\% writes. On the other hand, with file models, a large fraction of lookups benefit from the models and thus file learning performs better than the baseline. For read-heavy mixed workload (5\%), although level learning has benefits over the baseline, it performs worse than file learning for the same reasons above. 

Level learning can be beneficial for read-only settings: as shown in the table, level learning provides 10\% improvements over file learning. Thus, deployments that have only read-only workloads can benefit from level learning. Given that \sysname's goal is to provide faster lookups while supporting writes, levels are not an appropriate choice of granularity for learning. Thus, \sysname\ uses file learning by default. However, \sysname\ supports level learning as a configuration option that can be useful in read-only scenarios. %This is how \sysname\ uses our last learning guideline described in the previous section.  

\subsection{Cost vs. Benefit Analyzer}
\label{ssec-cba}

Before learning a file, \sysname\ must ensure that the time spent in learning is worthwhile. If a file is short-lived, then the time spent learning that file wastes resources. Such a file will serve few lookups and thus the model would have little benefit. Thus, to decide whether or not to learn a file, \sysname\ implements an online cost vs. benefit analysis. 

\subsubsection{Wait Before Learning}

As our analysis showed, even in the lower levels, many files are short-lived. To avoid the cost of learning short-lived files, \sysname\ waits for a time threshold, $T_{wait}$, before learning a file. The exact value of $T_{wait}$ presents a cost vs. performance tradeoff. A very low $T_{wait}$ leads to some short-lived files still being learned, incurring overheads; a large value causes many lookups to take the baseline path (because there is no model built yet), thus missing opportunities to make lookups faster. \sysname\ sets the value of $T_{wait}$ to the time it takes to learn a file. Our approach is never more than a factor of two worse than the optimal solution, where the optimal solution knows apriori the lifetime and decides to either immediately or never learn the file (i.e., it is two-competitive~\cite{karlin1991empirical}). Through measurements, we found that the maximum time to learn a file \camera{(which is at most $\sim$4MB in size)} is around 40 ms \camera{on our experimental setup}. We conservatively set $T_{wait}$ to be 50 ms in \sysname's implementation. 

%Workload and Data Awareness
\subsubsection{To Learn a File or Not}

\sysname\ waits for $T_{wait}$ before learning a file. However, learning a file even if it lives for a long time may not be beneficial. For example, our analysis shows that although lower-level files live longer, for some workloads and datasets, they serve relatively fewer lookups than higher-level files; higher-level files, although short-lived, serve a large percentage of negative internal lookups in some scenarios. \sysname, thus, must consider the potential benefits that a model can bring, in addition to considering the cost to build the model. It is profitable to learn a file if the benefit of the model ($B_{model}$) outweighs the cost to build the model ($C_{model}$). 

\noindent
\textbf{\textit{Estimating $\mathbf{C_{model}}$.}} One way to estimate $C_{model}$ is to assume that the learning is completely performed in the background and will not affect the rest of the system; i.e., $C_{model}$ is 0. This is true if there are many idle cores which the learning threads can utilize and thus do not interfere with the foreground tasks (e.g., the workload) or other background tasks (e.g., compaction). However, \sysname\ takes a conservative approach and assumes that the learning threads will interfere and slow down the other parts of the system. As a result, \sysname\ assumes $C_{model}$ to be equal to $T_{build}$. We define $T_{build}$ as the time to train the PLR model for a file. We find that this time is linearly proportional to the number of data points in the file. We calculate $T_{build}$ for a file by multiplying the average time to a train a data point (measured offline) and the number of data points in the file.

\noindent
\textbf{\textit{Estimating $\mathbf{B_{model}}$.}} Estimating the potential benefit of learning a file, $B_{model}$, is more involved. Intuitively, the benefit offered by the model for an internal lookup is given by $T_{b} - T_{m}$, where $T_{b}$ and $T_{m}$ are the average times for the lookup in baseline and model paths, respectively. If the file serves N lookups in its lifetime, the net benefit of the model is: $B_{model} = (T_{b} - T_{m}) * N$. We divide the internal lookups into negative and positive because most negative lookups terminate at the \camera{filter}, whereas positive ones do not; thus, 

\vspace{-0.25in}
\begin{gather*}
B_{model} = ((T_{n.b} - T_{n.m}) * N_{n}) + ((T_{p.b} - T_{p.m}) * N_{p})
\end{gather*}
\vspace{-0.25in}

%the number of negative ($N_{n}$) and positive ($N_{p}$) internal lookups

\noindent
where $N_{n}$ and $N_{p}$ are the number of negative and positive internal lookups, respectively. $T_{n.b}$ and $T_{p.b}$ are the time in the baseline path for a negative and a positive lookup, respectively; $T_{n.m}$ and $T_{p.m}$ are the model counterparts.% lookup times for a negative and a positive internal lookup, respectively.

$B_{model}$ for a file cannot be calculated without knowing the number of lookups that the file will serve or how much time the lookups will take. The analyzer, to  estimate these quantities, maintains statistics of files that have lived their lifetime, i.e., files that were created, served many lookups, and then were replaced. To estimate these quantities for a file $F$, the analyzer uses the statistics of other files at the same level as $F$; we consider statistics only at the same level because these statistics vary significantly across levels.        

Recall that \sysname\ waits before learning a file. During this time, the lookups are served in the baseline path. \sysname\ uses the time taken for these lookups to estimate $T_{n.b}$ and $T_{p.b}$. Next, $T_{n.m}$ and $T_{p.m}$ are estimated as the average negative and positive model lookup times of other files at the same level. Finally, $N_{n}$ and $N_{p}$ are estimated as follows. The analyzer first takes the average negative and positive lookups for other files in that level; then, it is scaled by a factor $f = s/\bar{s_l}$, where $s$ if the size of the file and $\bar{s_l}$ is the average file size at this level. While estimating the above quantities, \sysname\ filters out very short-lived files. 

While bootstrapping, the analyzer might not have enough statistics collected. Therefore, initially, \sysname\ runs in an always-learn mode (with $T_{wait}$ still in place.) Once enough statistics are collected, the analyzer performs the cost vs. benefit analysis and chooses to learn a file if $C_{model} < B_{model}$, i.e., benefit of a model outweighs the cost. If multiple files are chosen to be learned at the same time, \sysname\ puts them in a max priority queue ordered by $B_{model} - C_{model}$, thus prioritizing files that would deliver the most benefit. 

\camera{Our cost-benefit analyzer adopts a simple scheme of using average statistics of other files at the same level. While this approach has worked well in our initial prototype, using more sophisticated statistics and considering workload distributions (e.g., to account for keys with different popularity) could be more beneficial. We leave such exploration for future work.}

\subsection{Bourbon: Putting it All Together}

We describe how the different pieces of \sysname\ work together. Figure~\ref{fig-llsm} shows the path of lookups in \sysname. As shown in (a), lookups can either be processed via the model (if the target file is already learned), or in the baseline path (if the model is not built yet.) The baseline path in \sysname\ is similar to the one shown in Figure~\ref{fig-leveldb} for LevelDB, except that \sysname\ stores the values separately and so {\em ReadValue} reads the value from the log. 

\input{fig-llsm}

Once \sysname\ learns a sstable file, lookups to that file will be processed via the learned model as shown in Figure~\ref{fig-llsm}(b). \circled{1} {\em FindFiles}: \sysname\ finds the candidate sstables; this step required because \sysname\ uses file learning. \circled{2} {\em LoadIB+FB}: \sysname\ loads the index and filter blocks; these blocks are likely to be already cached. \circled{3} {\em ModelLookup}: \sysname\ performs a look up for the desired key $k$ in the candidate sstable's model. The model outputs a predicted position of $k$ within the file ($pos$) and the error bound ($\delta$). From this, \sysname\ calculates the data block that contains records $pos - \delta$ through $pos + \delta$.\footnote[2]{Sometimes, records $pos - \delta$ through $pos + \delta$ span multiple data blocks; in such cases, \sysname\ consults the index block (which specifies the maximum key in each data block) to find the data block for $pos$.} \circled{4} {\em SearchFB}: The filter for that block is queried to check if $k$ is present. If present, \sysname\ calculates the range of bytes of the block that must be loaded; this is simple because keys and pointers to values are of fixed size. \circled{5} {\em LoadChunk}: The byte range is loaded. \circled{6} {\em LocateKey}: The key is located in the loaded chunk. The key will likely be present in the predicted position (the midpoint of the loaded chunk); if not, \sysname\ performs a binary search in the chunk. \circled{7} {\em ReadValue}: The value is read from the value log using the pointer.

\noindent
\textbf{Possible improvements.} Although \sysname's implementation is highly-optimized and provides many features common to real systems, it lacks a few features. For example, in the current implementation, we do not support string keys and key compression (although we support value compression). \camera{For string keys, one approach we plan to explore is to treat strings as base-64 integers and convert them into 64-bit integers, which could then adopt the same learning approach described herein. While this approach may work well for small keys, large keys may require larger integers (with more than 64 bits) and thus efficient large-integer math is likely essential.} Also, \sysname\ does not support adaptive switching between level and file models; it is a static configuration. We leave supporting these features to future work.

%% file: tbl-granularity.tex
\setlength{\tempa}{\tabcolsep}
\newcolumntype{C}[1]{>{\centering\arraybackslash}p{#1}}
\newcolumntype{M}[1]{>{\centering\arraybackslash}m{#1}}
\setlength{\tabcolsep}{1.1pt}
\begin{table}[!t]
  \centering
{\footnotesize
\begin{tabular} {M{1.5cm}|M{1cm}|M{1cm}|M{1.4cm}|M{1cm}|M{1.4cm}}
\multirow{2}{1.5cm}{Workload} 			 & \multirow{2}{1cm}{Baseline time (s)} 	& \multicolumn{2}{c|}{File model} 			&  \multicolumn{2}{c}{Level model} 	 \\\cline{3-6}
									 &	& Time(s) & \% model & Time(s) & \% model \\\hline

Mixed: Write-heavy &	82.6  			&	71.5 (1.16 $\times$)  & 74.2	& 	95.1 (0.87 $\times$) & 1.5 \\\hline
Mixed: Read-heavy   &	89.2  			&	62.05 (1.44 $\times$)	& 99.8  &	74.3 (1.2 $\times$) & 21.4 \\\hline
Read-only	 &	48.4  			&	27.2 (1.78 $\times$) & 100 	& 	25.2 (1.92 $\times$) & 100 \\

\end{tabular}  
}
%\vspace{-0.05in}
\mycaption{tbl-granularity}{File vs. Level Learning}{The table compares the time to perform 10M operations in baseline WiscKey, file-learning, and level-learning. The numbers within the parentheses show the improvements over baseline. The table also shows the percentage of lookups that take the model path; remaining take the original path because the models are not rebuilt yet.}

\end{table}
\setlength{\tabcolsep}{\tempa}

%% file: fig-llsm.tex
\begin{figure}[t!]
\centering
\includegraphics[scale=0.86]{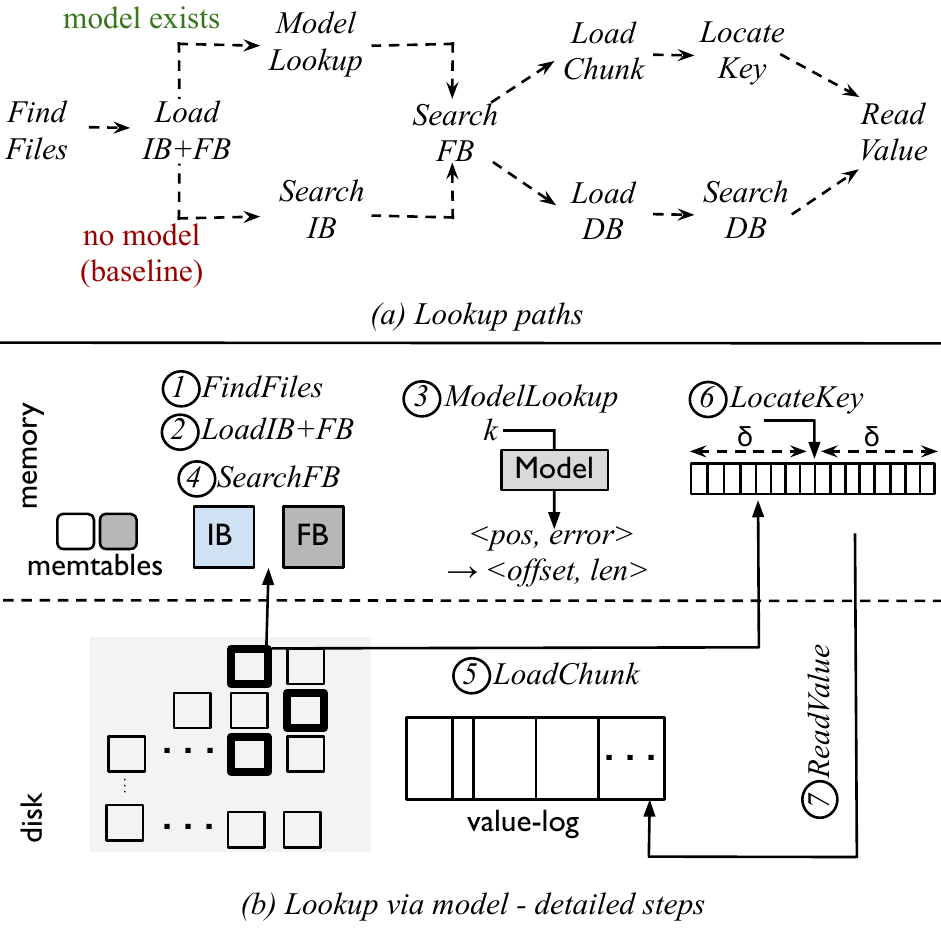}
\vspace{-0.05in}
\mycaption{fig-llsm}{\sysname\ Lookups}{(a) shows that lookups can take two different paths: when the model is available (shown at the top), and when the model is not learned yet and so lookups take the baseline path (bottom); some steps are common to both paths. (b) shows the detailed steps for a lookup via a model; we show the case where models are built for files.}
\end{figure}

%% file: eval.tex
\section{Evaluation}
\label{sec-eval}

To evaluate \sysname, we ask the following questions:
\begin{itemize}[noitemsep,nolistsep,topsep=0pt,parsep=0pt,partopsep=0pt,leftmargin=*]
  \setlength\itemsep{0em}
  \item Which portions of lookup does \sysname\ optimize? (\sref{whichportions})
  \item How does \sysname\ perform with models available and no writes? How does performance change with datasets, load orders, and request distributions? (\sref{nowrites})
  \item \camera{How does \sysname\ perform with range queries? (\sref{range})}
  \item In the presence of writes, how does \sysname's cost-benefit analyzer perform compared to other approaches that always or never re-learn? (\sref{writescba})
  \item Does \sysname\ perform well on real benchmarks? (\sref{realmacro})
  \item Is \sysname\ beneficial when data is on storage? (\sref{optanessd})
  \item \camera{Is \sysname\ beneficial with limited memory?
  (\sref{limitedmem})}
  \item What are the error and space tradeoffs of \sysname? (\sref{modelerror})
\end{itemize}

\input{fig-datasets}

\noindent
\textbf{Setup.} We run our experiments on a 20-core Intel Xeon CPU E5-2660 machine with 160-GB memory and a 480-GB SATA SSD. We use 16B integer keys and 64B values, and set the error bound of \sysname's PLR as 8. Unless specified, our workloads perform 10M operations. We use a variety of datasets. We construct four synthetic datasets: linear, segmented-1\%, segmented-10\% , and normal, each with 64M key-value pairs. In the linear dataset, keys are all consecutive. In the seg-1\% dataset, there is a gap after a consecutive segment of 100 keys (i.e., every 1\% causes a new segment). The segmented-10\% dataset is similar, but there is a gap after 10 consecutive keys. We generate the normal dataset by sampling 64M unique values from the standard normal distribution $N(0, 1)$ and scale to integers. We also use two real-world datasets: Amazon reviews (AR)~\cite{ardataset} and New York OpenStreetMaps (OSM)~\cite{osmdataset}. AR and OSM have 33.5M and 21.9M key-value pairs, respectively. These datasets vary widely in how the keys are distributed. Figure~\ref{fig-datasets} shows the distribution for a few datasets. Most of our experiments focus on the case where the data resides in memory; however, we also analyze cases where data is present on storage. 
%of the keys and their positions

\subsection{Which Portions does \sysname\ Optimize?}
\label{whichportions}

We first analyze which portions of the lookup \sysname\ optimizes. We perform 10M random lookups on the AR and OSM datasets and show the latency breakdown in Figure~\ref{fig-splitup}. As expected, \sysname\ reduces the time spent in indexing. The portion marked {\em Search} in the figure corresponds to {\em SearchIB} and {\em SearchDB} in the baseline, versus {\em ModelLookup} and {\em LocateKey} in \sysname. The steps in \sysname\ have lower latency than their baseline counterparts. Interestingly, \sysname\ reduces data-access costs too, because \sysname\ loads a smaller byte range than the entire block loaded by the baseline.

\input{fig-splitup}

\subsection{Performance under No Writes}
\label{nowrites}

We next analyze \sysname's performance when the models are already built and there are no updates. For each experiment, we load a dataset and allow the system to build the models; during the workload, we issue only lookups. 

\input{fig-datasets-perf}

\subsubsection{Datasets}

To analyze how the performance is influenced by the dataset, we run the workload on all six datasets and compare \sysname's lookup performance against WiscKey. Figure~\ref{fig-micro-datasets} show the results. As shown in ~\ref{fig-micro-datasets}(a), \sysname\ is faster than WiscKey for all datasets; depending upon the dataset, the improvements vary (1.23$\times$ to 1.78$\times$). \sysname\ provides the most benefit for the linear dataset because it has the smallest number of segments (one per model); with fewer segments, fewer searches are needed to find the target line segment. From ~\ref{fig-micro-datasets}(b), we observe that latencies increase with the number of segments (e.g., latency of seg-1\% is greater than that of linear). We cannot compare the number of segments in AR and OSM with others because the size of these datasets is significantly different.

\noindent
\textbf{Level learning}. Given that level learning is suitable for read-only scenarios, we configure \sysname\ to use level learning and analyze its performance. As shown in Figure~\ref{fig-micro-datasets}(a), \sysname-level is 1.33$\times$ -- 1.92$\times$ faster than the baseline. \sysname-level offers more benefits than \sysname\ because a level-model lookup is faster than finding the candidate sstables and then doing a file-model lookup. This confirms that \sysname-level is an attractive option for read-only scenarios. However, since level models only provide benefits for read-only workloads and give at most 10\% improvement compared to file models, we focus on \sysname\ with file learning for our remaining experiments.  

\input{fig-arosm}

\subsubsection{Load Orders}

We now explore how the order in which the data is loaded affects performance. 
For this experiment, we use the AR and OSM datasets and load them in two ways: sequential (keys are inserted in ascending order) and random (keys are inserted in an uniformly random order). With sequential loading, sstables do not have overlapping key ranges even across levels; whereas, with random loading, sstables at one level can overlap with sstables at other levels.

Figure~\ref{fig-arosm} shows the result. First, regardless of the load order, \sysname\ offers significant benefit over baseline (1.47$\times$ -- 1.61$\times$). Second, the average lookup latencies increase in the randomly-loaded case compared to the sequential case (e.g., 6$\mu$s vs. 4$\mu$s in WiscKey for the AR dataset). This is because while there are no negative internal lookups in the sequential case, there are many (23M) negative lookups in the random case (as shown in ~\ref{fig-arosm}(b)). Thus, with random load, the total number of internal lookups increases by 3$\times$, increasing lookup latencies. 

Next, we note that the speedup over baseline in the random case is less than that of the sequential case (e.g., 1.47$\times$ vs. 1.61$\times$ for AR). Although \sysname\ optimizes both positive and negative internal lookups, the gain for negative lookups is smaller (as shown in ~\ref{fig-arosm}(b)). This is because most negative lookups in the baseline and \sysname\ end just after the filter is queried (filter indicates absence); the data block is not loaded or searched. Given there are more negative than positive lookups, \sysname\ offers less speedup than the sequential case. However, this speedup is still significant (1.47$\times$).

\input{fig-reqdist}

\subsubsection{Request Distributions}

Next, we analyze how request distributions affect \sysname's performance. We measure the lookup latencies under six request distributions: sequential, zipfian, hotspot, exponential, uniform, and latest. We first randomly load the AR and OSM datasets and then run the workloads; thus, the data can be segmented and there can be many negative internal lookups. As shown in Figure~\ref{fig-reqdist}, \sysname\ makes lookups faster by 1.54$\times$ -- 1.76$\times$ than the baseline. Overall, \sysname\ reduces latencies regardless of request distributions.

\noindent
\textbf{Read-only performance summary.} When the models are already built and when there are no writes, \sysname\ provides significant speedup over baseline for a variety of datasets, load orders, and request distributions.

\input{fig-range}
\input{fig-cba-perf}
\input{fig-macro-ycsb}

\subsection{Range Queries}
\label{range}
\camera{We next analyze how \sysname\ performs on range queries. We perform 1M range queries on the AR and OSM datasets with various range lengths. Figure~\ref{fig-range} shows the throughput of \sysname\ normalized to that of WiscKey. With short ranges, where the indexing  cost (i.e., the cost to locate the first key of the range) is dominant, \sysname\ offers the most benefit. For example, with a range length of 1 on the AR dataset, \sysname\ is 1.90$\times$ faster than WiscKey. The gains drop as the range length increases; for example, \sysname\ is only 1.15$\times$ faster with queries that return 100 items. This is because, while \sysname\ can accelerate the indexing portion, it follows a similar path as WiscKey to scan subsequent keys. Thus, with large range lengths, indexing accounts for less of the total performance, resulting in lower gains.}

\subsection{Efficacy of Cost-benefit Analyzer with Writes}
\label{writescba}

We next analyze how \sysname\ performs in the presence of writes. Writes modify the data and so the models must be re-learned. In such cases, the efficacy of \sysname's cost-benefit analyzer (cba) is critical. We thus compare \sysname's cba against two strategies: \sysname-offline and \sysname-always. \sysname-offline performs no learning as writes happen; models exist only for the initially loaded data. \sysname-always re-learns the data as writes happen; it always decides to learn a file without considering cost. \sysname-cba re-learns as well, but it uses the cost-benefit analysis to decide whether or not to learn a file. 

We run a workload that issues 50M operations with varying percentages of writes on the AR dataset. To calculate the total amount of work performed for each workload, we sum together the time spent on the foreground lookups and inserts (Figure~\ref{fig-cba-perf}(a)), the time spent learning (\ref{fig-cba-perf}(b)), and the time spent on compaction (not shown); the total amount of work is shown in Figure \ref{fig-cba-perf}(c). The figure also shows the fraction of internal lookups that take the baseline path (\ref{fig-cba-perf}(d)).

First, as shown in \ref{fig-cba-perf}(a), all \sysname\ variants reduce the workload time compared to WiscKey. The gains are lower with more writes because \sysname\ has fewer lookups to optimize. Next, \sysname-offline performs worse than \sysname-always and \sysname-cba. Even with just 1\% writes, a significant fraction of internal lookups take the baseline path in \sysname-offline as shown in \ref{fig-cba-perf}(d); this shows re-learning as data changes is crucial.

\sysname-always learns aggressively and thus almost no lookups take the baseline path even for 50\% writes. As a result, \sysname-always has the lowest foreground time. However, this comes at the cost of increased learning time; for example, with 50\% writes, \sysname-always spends about 134 seconds learning. Thus, the total time spent increases with more writes for \sysname-always and is even higher than baseline WiscKey as shown in \ref{fig-cba-perf}(c). Thus, aggressively learning is not ideal.

Given a low percentage of writes, \sysname-cba decides to learn almost all the files, and thus matches the characteristics of \sysname-always: both have a similar fraction of lookups taking the baseline path, both require the same time learning, and both perform the same amount of work. With a high percentage of writes, \sysname-cba chooses not to learn many files, reducing learning time; for example, with 50\% writes, \sysname-cba spends only 13.9 seconds in learning (10$\times$ lower than \sysname-always). Consequently, many lookups take the baseline path. \sysname-cba takes this action because there is less benefit to learning as the data is changing rapidly and there are fewer lookups. Thus, it almost matches the foreground time of \sysname-always. But, by avoiding learning, the total work done by \sysname-cba is significantly lower. 

\noindent
\textbf{Summary.} Aggressive learning offers fast lookups but with high costs; no re-learning provides little speedup. Neither is ideal. In contrast, \sysname\ provides high benefits similar to aggressive learning while lowering total cost significantly.

\subsection{Real Macrobenchmarks}
\label{realmacro}

We next analyze how \sysname\ performs under two real benchmarks: YCSB~\cite{Cooper:2010:BCS:1807128.1807152} and SOSD~\cite{kipf2019sosd}.

\subsubsection{YCSB}

We use \camera{six} workloads that have different read-write ratios and access patterns: A (w:50\%, r:50\%), B (w:5\%, r:95\%), C (read-only), D (read latest, w:5\%, r:95\%), \camera{E (range-heavy, w:5\%, range:95\%),} F (read-modify-write:50\%, r:50\%). We use three datasets: YCSB's default dataset (created using {\em ycsb-load}~\cite{ycsb-load}), AR, and OSM, and load them in a random order. Figure~\ref{fig-macro-ycsb} shows the results.

For the read-only workload (YCSB-C), all operations benefit and \sysname\ offers the most gains (about 1.6$\times$). For read-heavy workloads (YCSB-B and D), most operations benefit, while writes are not improved and thus \sysname\ is 1.24$\times$ -- 1.44$\times$ faster than the baseline. For write-heavy workloads (YCSB-A and F), \sysname\ improves performance only a little (1.06$\times$ -- 1.18$\times$). First, a large fraction of operations are writes; second, the number of the internal lookups taking the model path decreases (by about 30\% compared to the read-heavy workload because \sysname\ chooses not to learn some files). \camera{YCSB-E consists of range queries (range lengths varying from 1 to 100) and 5\% writes. \sysname\ reaches 1.16$\times$ -- 1.19$\times$ gain.} In summary, as expected, \sysname\ improves the performance of read operations; at the same time, \sysname\ does not affect the performance of writes.

\subsubsection{SOSD}

We next measure \sysname's performance on the SOSD benchmark designed for learned indexes~\cite{kipf2019sosd}. We use the following six datasets: book sale popularity (amzn32), Facebook user ids (face32), lognormally (logn32) and normally (norm32) distributed datasets, uniformly distributed dense (uden32) and sparse (uspr32) integers. Figure~\ref{fig-macro-sosd} shows the average lookup latency. As shown, \sysname\ is about 1.48$\times$ -- 1.74$\times$ faster than the baseline for all datasets. 

\input{fig-macro-sosd}
\input{tbl-optane}
\input{fig-optane-ycsb}

\subsection{Performance on Fast Storage}
\label{optanessd}

Our analyses so far focused on the case where the data resides in memory. We now analyze if \sysname\ will offer benefit when the data resides on a fast storage device. We run a read-only workload on sequentially loaded AR and OSM datasets on an Intel Optane SSD. \camera{Table~\ref{tbl-optane}} shows the result. Even when the data is present on a storage device, \sysname\ offers benefit (1.25$\times$ -- 1.28$\times$ faster lookups than WiscKey). \camera{Figure~\ref{fig-optane-ycsb} shows the result for read-write mixed YCSB workloads on the same device with the default YCSB datasest. As expected, while \sysname's benefits are marginal for write-heavy workloads (YCSB-A and YCSB-F), it offers considerable speedup (1.16$\times$ -- 1.19$\times$) for read-heavy workloads (YCSB-B and YCSB-D).} With the emerging storage technologies (e.g., 3D XPoint memory), \sysname\ will offer even more benefits.

\subsection{Performance with Limited Memory}
\label{limitedmem}
\input{tbl-mem}

\camera{We further show that, even with no fast storage and limited available memory, \sysname\ can still offer benefit with skewed workloads, such as zipfian. We experiment on a machine with a SATA SSD and memory that only holds about 25\% of the database. We run a uniform random workload, and a zipfian workload with consecutive hotspots where 80\% of the requests access about 25\% of the database. Table~\ref{tbl-mem} shows the result. With the uniform workload, \sysname\ is only 1.04$\times$ faster because most of the time is spent loading the data into the memory. With the zipfian workload, in contrast, indexing time instead of data-access time dominates because a large number of requests access the small portion of data that is already cached in memory. \sysname\ is able to reduce this significant indexing time and thus offers 1.25$\times$ lower latencies.}

\subsection{Error Bound and Space Overheads}
\label{modelerror}

We finally discuss the characteristics of \sysname's ML model, specifically its error bound ($\delta$) and space overheads. Figure~\ref{fig-error}(a) plots the error bound ($\delta$) against the average lookup latency (left y-axis) for AR dataset. As $\delta$ increases, fewer line segments are created, leading to fewer searches, thus reducing latency. However, beyond $\delta = 8$, although the time to find the segment reduces, the time to search within a segment increases, \camera{thus} increasing latency. We find that \sysname's choice of $\delta = 8$ is optimal for other datasets too. \camera{Figure~}\ref{fig-error}(a) also shows how space overheads (right y-axis) vary with $\delta$. As $\delta$ increases, fewer line segments are created, leading to low space overheads. Table~\ref{fig-error}(b) shows the space overheads for different datasets. As shown, for a variety of datasets, the overhead compared to the total dataset size is little (0\% -- 2\%).

%% file: fig-datasets.tex
\begin{figure}[t!] 
\begin{center}
\vspace{-0.15in}
\subfloat[Linear]
{
\includegraphics[scale=0.41]{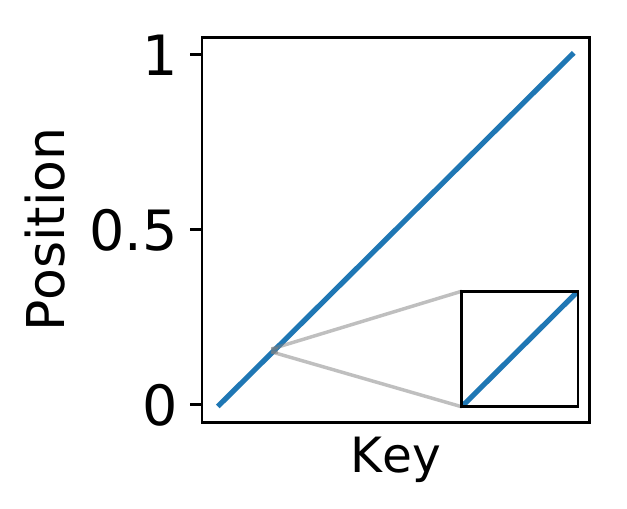}
}
\hspace{-0.16in}
\subfloat[Seg10\%]
{
\includegraphics[scale=0.41]{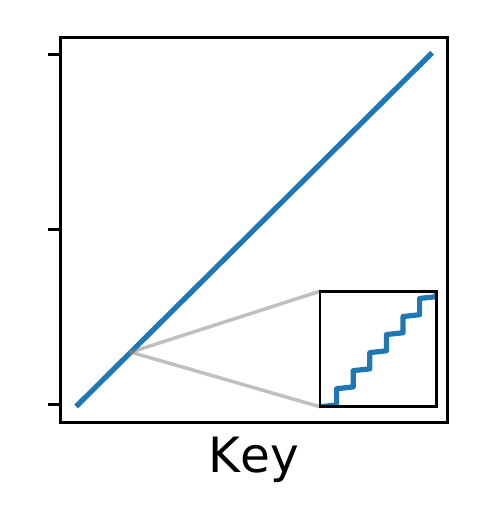}
}
\hspace{-0.16in}
\subfloat[Normal]
{
\includegraphics[scale=0.41]{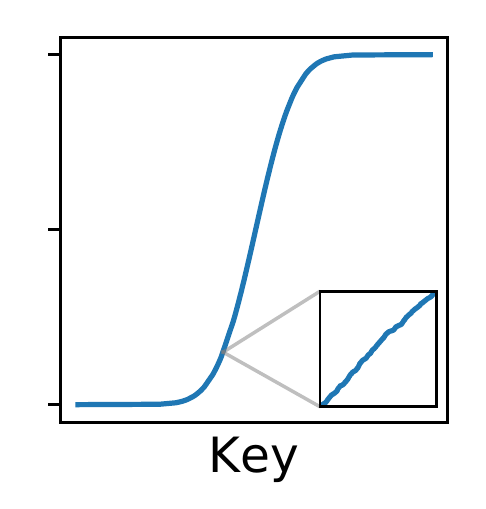}
}
\hspace{-0.16in}
\subfloat[OSM]
{
\includegraphics[scale=0.41]{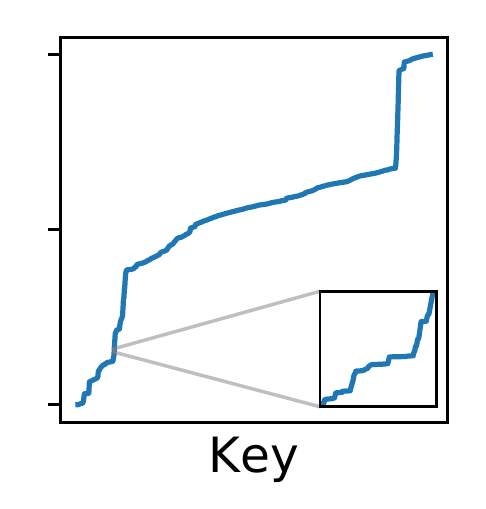}
}
\vspace{-0.1in}
\end{center}
\mycaption{fig-datasets}{Datasets}{The figure shows the cumulative distribution functions (CDF) of three synthetic datasets (linear, segmented-10\%, and normal) and one real-world dataset (OpenStreetMaps). Each dataset is magnified around the 15\% percentile to show a detailed view of its distribution.}
\end{figure}

%\subfloat[Seg1\%]
%{
%\includegraphics[scale=0.4]{figs/ds-Seg11.pdf}
%}

%\subfloat[AR]
%{
%\includegraphics[scale=0.4]{figs/ds-AR1.pdf}
%}
%The figure shows the cumulative distribution functions (CDF) of the four synthetic datasets (linear, segmented-1\%, segmented-10\%, and normal) and the two real-world datasets (Amazon Reviews and OpenStreetMaps). Each dataset is magnified around the 15\% percentile to show a detailed view of its distribution.

%% file: fig-splitup.tex
\begin{figure}[t!]
\centering
\includegraphics[scale=1.1]{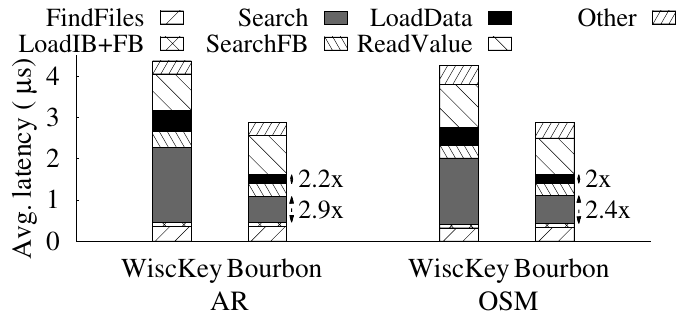}
\vspace{-0.05in}
\mycaption{fig-splitup}{Latency Breakdown}{The figure shows latency breakdown for WiscKey and \sysnamesmall. Search denotes SearchIB and SearchDB in WiscKey; the same denotes ModelLookup and LocateKey in \sysnamesmall. LoadData denotes LoadDB in WiscKey; the same denotes LoadChunk in \sysnamesmall. These two steps are optimized by \sysnamesmall\ and are shown in solid colors; the number next to a step shows the factor by which it is made faster in \sysnamesmall.}
\end{figure}

%; others steps are shown in patterns

%% file: fig-datasets-perf.tex
\begin{figure}[t!] 
\captionsetup[subfigure]{labelformat=empty, justification=centering}
\subfloat[(a) Average lookup latency]
{
\begin{minipage}{0.295\textwidth}\
\hspace{-0.1in} \includegraphics[scale=0.46]{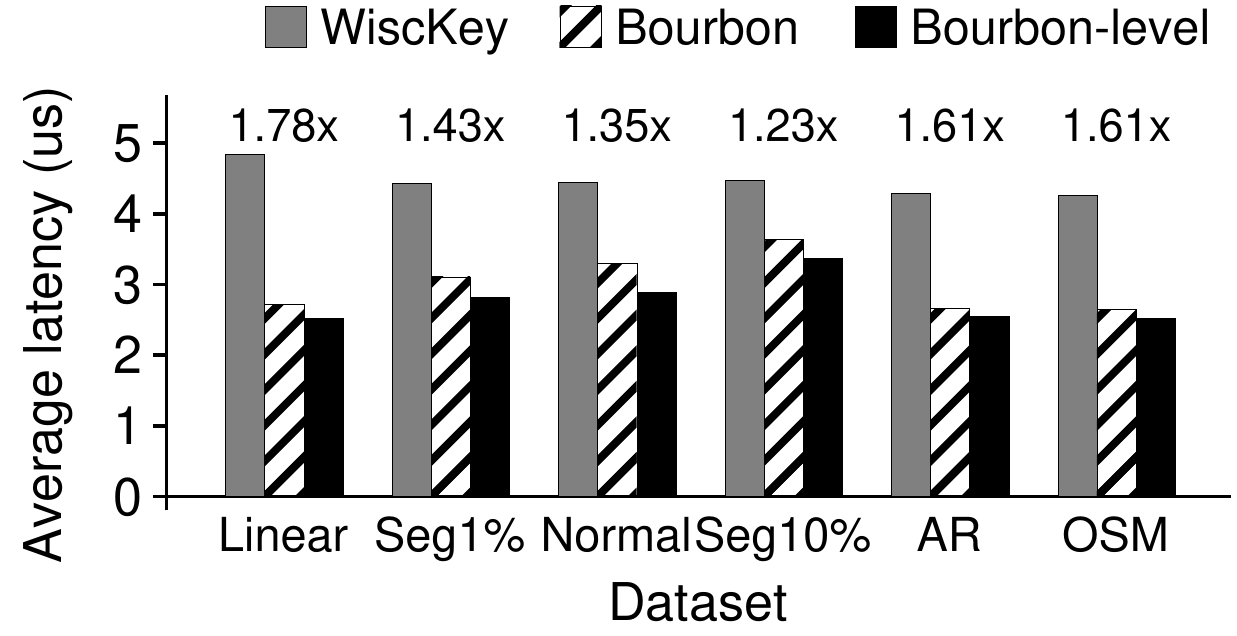}
\end{minipage}%
}
\subfloat[(b) Number of segments]
{
\begin{minipage}{0.185\textwidth}\
\input{tbl-dataset}
\end{minipage}%
}
\vspace{-0.03in}
\mycaption{fig-micro-datasets}{Datasets}{(a) compares the average lookup latencies of \sysnamesmall, \sysnamesmall-level, and WiscKey for different datasets; the numbers on the top show the improvements of \sysnamesmall\ over WiscKey. (b) shows the number of segments for different datasets in \sysnamesmall.}
\end{figure}

%% file: tbl-dataset.tex
\setlength{\tempa}{\tabcolsep}
\newcolumntype{C}[1]{>{\centering\arraybackslash}p{#1}}
\newcolumntype{M}[1]{>{\centering\arraybackslash}m{#1}}
\setlength{\tabcolsep}{1.05pt}
{\footnotesize
\begin{tabular} {M{0.9cm}|M{0.7cm}|M{0.75cm}}
{Dataset} & \#segs & {latency ($\mu$s)} \\\hline
Linear	 &	900 &	2.72 \\
Seg1\%	 & 640K	&	3.11 \\
Normal	 & 705K	&	3.3	\\
Seg10\%	 & 6.4M	&	3.64 \\\hline
AR	 & 129K	&	2.66 \\\hline
OSM	 & 295K	&	2.65 \\
\end{tabular}  
}
%\mycaption{tbl-dataset}{Number of Segments in \sysnamesmall}{\footnotesize The table shows the number of segments for different datasets in \sysnamesmall.}

\setlength{\tabcolsep}{\tempa}

%% file: fig-arosm.tex
\begin{figure}[t!] 
\centering
\captionsetup[subfigure]{labelformat=empty, justification=centering}
\subfloat[(a) Average latency]
{
\begin{minipage}{0.21\textwidth}\
 \centering
\includegraphics[scale=0.35]{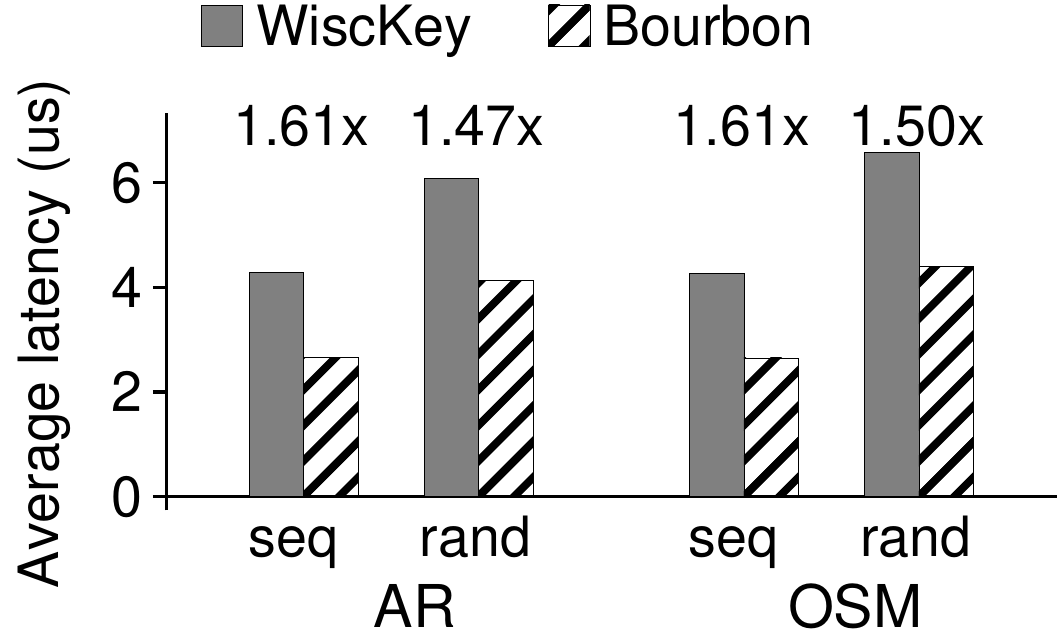}
\end{minipage}%
}
\subfloat[(b) Positive vs. negative internal lookups for randomly loaded case]
{
\begin{minipage}{0.26\textwidth}\
 \centering
\input{tbl-negpos}
\end{minipage}%
}

\vspace{-0.03in}
\mycaption{fig-arosm}{Load Orders}{(a) shows the performance for AR and OSM datasets for sequential (seq) and random (rand) load orders. (b) compares the speedup of positive and negative internal lookups.}

\end{figure}

%% file: tbl-negpos.tex
\setlength{\tempa}{\tabcolsep}
\newcolumntype{C}[1]{>{\centering\arraybackslash}p{#1}}
\newcolumntype{M}[1]{>{\centering\arraybackslash}m{#1}}
\setlength{\tabcolsep}{1pt}

{\footnotesize
\begin{tabular} {M{0.8cm}|M{0.6cm}|M{1cm}|M{0.6cm}|M{1cm}}
\multirow{2}{0.8cm}{Dataset} & \multicolumn{2}{M{1.6cm}|}{Positive} &  \multicolumn{2}{M{1.6cm}}{Negative} \\\cline{2-5}
									 & \# & Speedup & \# & Speedup \\\hline
									 %Sequential	NumSegments(File)	time	Gain-File		LLSM-File	
AR	 &	10M &	2.15$\times$ &  23M &	1.83$\times$ \\
OSM	 & 	10M	&	1.99$\times$ &  22M &	1.82$\times$ \\
\end{tabular}  
}
%\mycaption{tbl-micro}{Number of Segments and Negative lookups in \sysname-file}{\footnotesize The table shows the number of segments (for sequential loading) and negative lookups (for random loading) for different dataset in \sysname-file.}

\setlength{\tabcolsep}{\tempa}

%% file: fig-reqdist.tex
\begin{figure}[t!]
\centering
\includegraphics[scale=0.45]{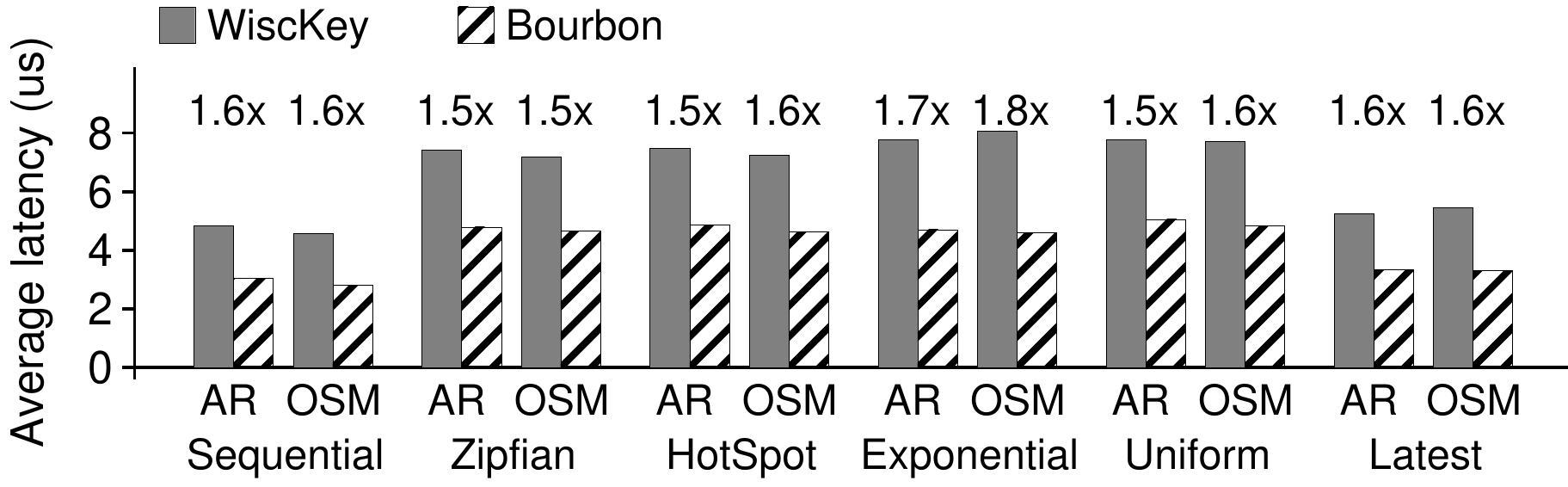}
\mycaption{fig-reqdist}{Request Distributions}{The figure shows the average lookup latencies of different request distributions from AR and OSM datasets.}
\end{figure}
%The numbers on the top show the performance normalized to that of the baseline

%% file: fig-range.tex
\begin{figure}[t!]
\centering
\includegraphics[scale=0.47]{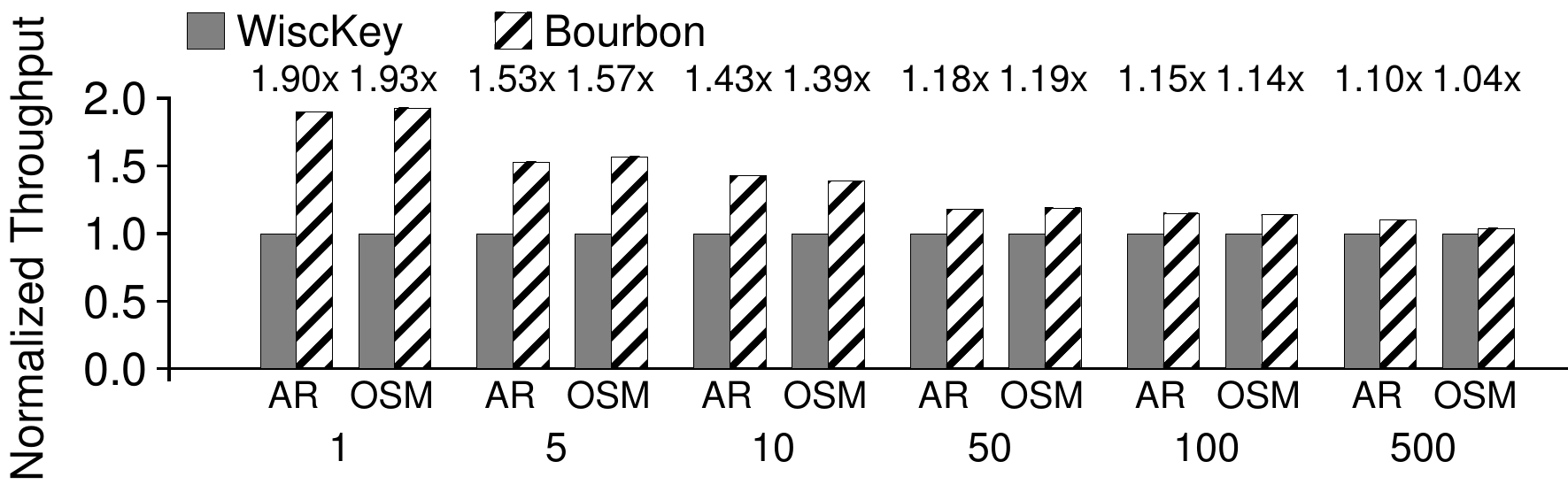}
\mycaption{fig-range}{Range Queries}{\camera{The figure shows the normalized throughput of range queries with different range lengths from AR and OSM datasets.}}
\end{figure}
%The numbers on the top show the performance normalized to that of the baseline

%% file: fig-cba-perf.tex
%\newcommand{\rulesep}{\unskip\ \vrule\ }
\begin{figure*}[t!] 
\begin{center}
    \includegraphics[scale=0.85]{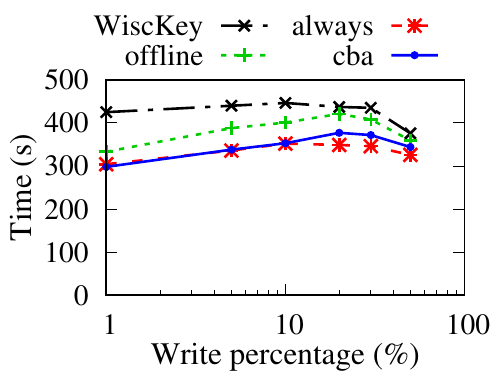}
    \includegraphics[scale=0.85]{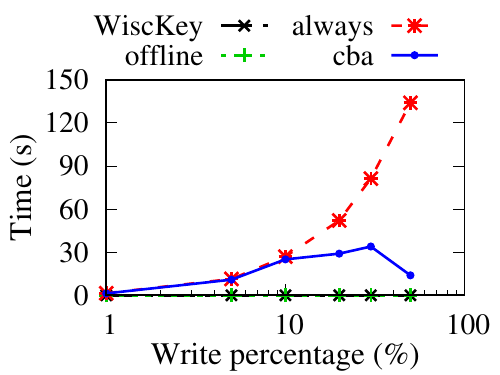}
    \includegraphics[scale=0.85]{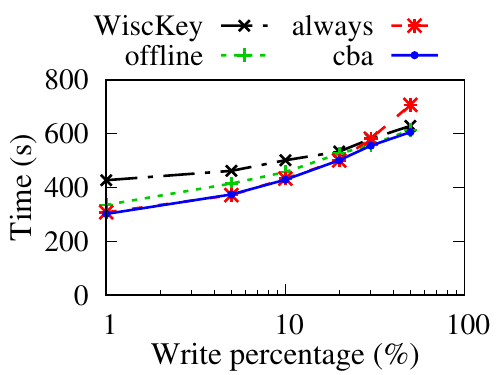}
	\includegraphics[scale=0.85]{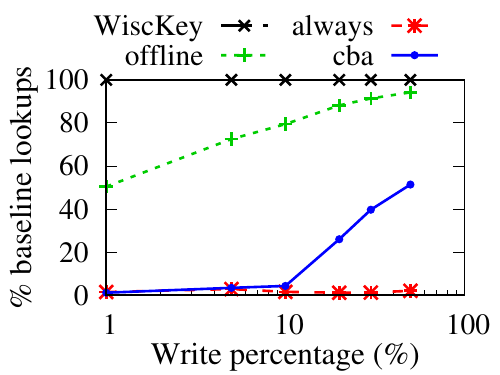}
\end{center}
\vspace{-0.2in}
\hspace{0.5in}{\footnotesize (a) Foreground time}\hspace{0.9in}{\footnotesize (b) Learning time}\hspace{1in}{\footnotesize (c) Total time}\hspace{0.6in}{\footnotesize (d) Baseline-path internal lookups (\%)}
%\hspace{-0.1in}{\footnotesize(a)(i) Synchronous Replication - Redis}\hspace{0.45in}{\footnotesize(a)(ii) Synchronous Replication - ZooKeeper}\hspace{0.45in}{\footnotesize(b) Asynchronous Replication - Redis}

%\hspace{0.5in}{\footnotesize(a) ZooKeeper}\hspace{1.3in}{\footnotesize(b) Redis (Synchronous Replication)}\hspace{0.45in}{\footnotesize(c) Redis (Asynchronous Replication)}

\vspace{-0.03in}
\mycaption{fig-cba-perf}{Mixed Workloads}{(a) compares the foreground times of WiscKey, \sysnamesmall-offline (offline), \sysnamesmall-always (always), and \sysnamesmall-cba (cba); (b) and (c) compare the learning time and total time, respectively; (d) shows the fraction of internal lookups that take the baseline path.}
\end{figure*}

%% file: fig-macro-ycsb.tex
\begin{figure*}[t!] 
\begin{center}
\includegraphics[scale=0.55]{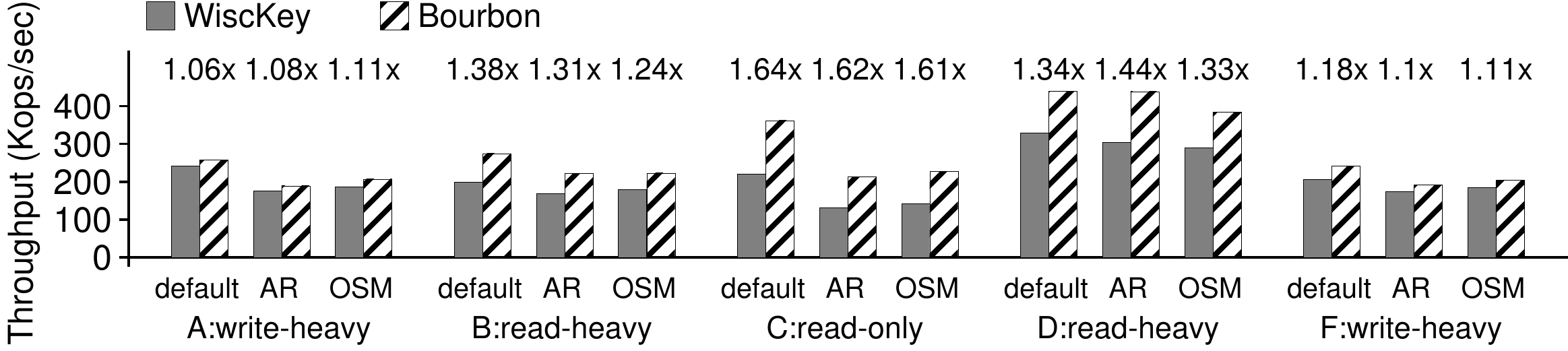}
%\hspace{0.15in}
\includegraphics[scale=0.55]{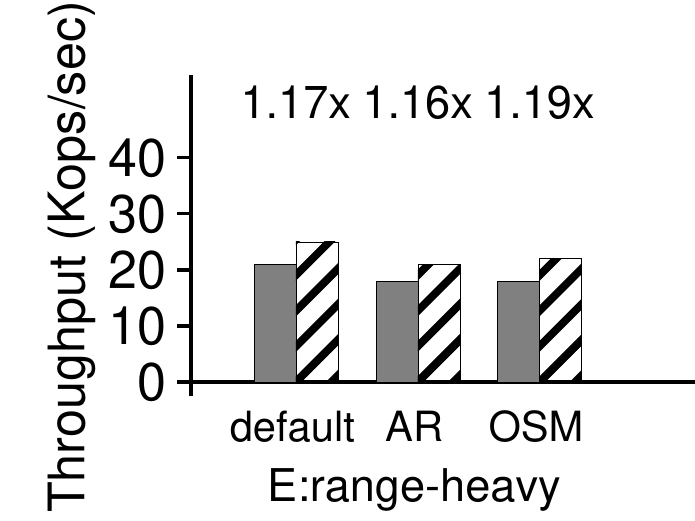}
% \footnotesize (a) YCSB

% \includegraphics[scale=0.46]{figs/sosd.eps}

% \footnotesize(b) SOSD
\end{center}
\vspace{-0.07in}
% \hspace{2in}{\footnotesize (a) YCSB}\hspace{3in}{\footnotesize(b) SOSD}
%\vspace{-0.03in}
\mycaption{fig-macro-ycsb}{Macrobenchmark-YCSB}{The figure compares the throughput of \sysnamesmall\ against WiscKey for \camera{six} YCSB workloads across three datasets.}
\vspace{0.10in}
\end{figure*}

%% file: fig-macro-sosd.tex
\begin{figure}[t!]
\centering
\includegraphics[scale=0.55]{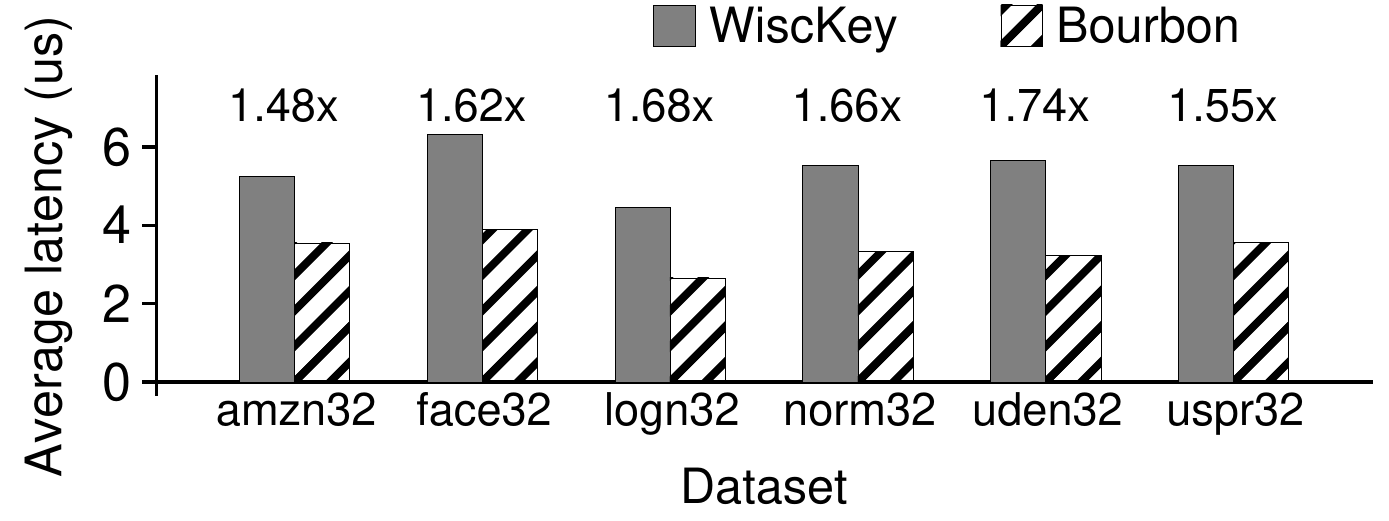}
\mycaption{fig-macro-sosd}{Macrobenchmark-SOSD}{\camera{The figure compares lookup latencies from the SOSD benchmark. The numbers on the top show \sysnamesmall's improvements over the baseline.}}
\end{figure}

%% file: tbl-optane.tex
\setlength{\tempa}{\tabcolsep}
\newcolumntype{C}[1]{>{\centering\arraybackslash}p{#1}}
\newcolumntype{M}[1]{>{\centering\arraybackslash}m{#1}}
\setlength{\tabcolsep}{1.1pt}
\begin{table}[!t]
    \centering
{\footnotesize
\begin{tabular} {c|M{1.5cm}|M{1.5cm}|M{1cm}}
\multirow{2}{*}{Dataset} & \multirow{2}{1.5cm}{\centering WiscKey latency ($\mu$s)} & \multicolumn{2}{c}{\sysnamesmall} \\\cline{3-4}
									& & Latency($\mu$s) & Speedup \\\hline

Amazon Reviews (AR) & 3.53	& 2.75 & 1.28$\times$ \\
NewYork OpenStreetMaps (OSM)	& 3.14	& 2.51 & 1.25$\times$ \\
\end{tabular}  
}
%\vspace{-0.02in}
\mycaption{tbl-optane}{Performance on Fast Storage}{The table shows \sysnamesmall's lookup latencies when the data is stored on an Optane SSD.}

\end{table}
\setlength{\tabcolsep}{\tempa}

%% file: fig-optane-ycsb.tex
\begin{figure}[t!]
\centering
\includegraphics[scale=0.55]{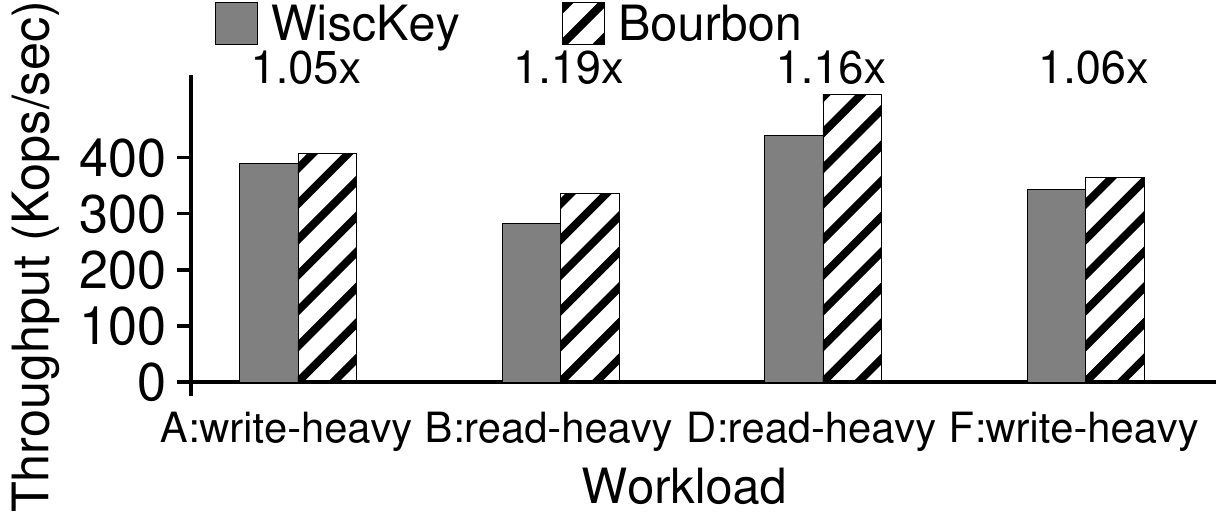}
\mycaption{fig-optane-ycsb}{Mixed Workloads on Fast Storage}{\camera{The figure compares the throughput of \sysname\ against WiscKey for four read-write mixed YCSB workloads. We use the YCSB default dataset for this experiment.}}
\end{figure}
%The numbers on the top show the performance normalized to that of the baseline

%% file: tbl-mem.tex
\setlength{\tempa}{\tabcolsep}
\newcolumntype{C}[1]{>{\centering\arraybackslash}p{#1}}
\newcolumntype{M}[1]{>{\centering\arraybackslash}m{#1}}
\setlength{\tabcolsep}{1.1pt}
\begin{table}[!t]
    \centering
{\footnotesize
\begin{tabular} {c|c|c|c}
\multirow{2}{*}{Workload} & \multirow{2}{1.5cm}{\centering WiscKey latency ($\mu$s)} & \multicolumn{2}{c}{\sysnamesmall} \\\cline{3-4}
									& & Latency($\mu$s) & Speedup \\\hline

Uniform & 98.6	& 94.4 & 1.04$\times$ \\
Zipfian	& 18.8	& 15.1 & 1.25$\times$ \\
\end{tabular}  
}
%\vspace{-0.05in}
\mycaption{tbl-mem}{Performance with Limited Memory}{\camera{The table shows \sysnamesmall's average lookup latencies from the AR dataset on a machine with a SATA SSD and limited memory.}}

\end{table}
\setlength{\tabcolsep}{\tempa}

%% file: new.related.tex
\section{Related Work}
\label{sec-related}
%Our work relates to and builds upon many prior efforts. %and builds upon 
% supporting updates: new data structures like gapped-array - our approach is complementary, models only for stable portions of the data -- modified b+ tree

% FITing tree -- modified b+ tree?

% high level: no prior work on learned indexes for LSM
%

\noindent
\textbf{Learned indexes.} The core idea of our work, replacing indexing structures with ML models, is inspired from the pioneering work on learned indexes~\cite{learnedindex}. However, learned indexes do not support updates, an essential operation that an storage-system index must support. Recent research tries to address this limitation. For instance, XIndex~\cite{tang2020xindex}, FITing-Tree~\cite{fittree}, and AIDEL~\cite{li2019scalable} support writes using an additional array (delta index) and with periodic re-training, whereas Alex~\cite{ding2019alex} uses gapped array at the leaf nodes of a B-tree to support writes. 
%and model-based inserts

\input{fig-error}

Most prior efforts optimize B- tree variants, while our work is the first to deeply focus on LSMs. Further, while most prior efforts implement learned indexes to stand-alone data structures, our work is the first to show how learning can be integrated and implemented into an existing, optimized, production-quality system. While SageDB~\cite{sagedb} is a full database system that uses learned components, it is built from scratch with learning in mind. Our work, in contrast, shows how learning can be integrated into an existing, practical system. Finally, instead of \quotes{fixing} new read-optimized learned index structures to handle writes (like previous work), we incorporate learning into an already write-optimized, production-quality LSM.

\vspace{0.1in}
\noindent
\textbf{LSM optimizations.} Prior work has built many LSM optimizations. Monkey~\cite{dayan2017monkey} carefully adjusts the bloom filter allocations for better filter hit rates and memory utilization. Dostoevsky~\cite{dayan2018dostoevsky}, HyperLevelDB~\cite{hyperleveldb}, and bLSM~\cite{sears2012blsm} develop optimized compaction policies to achieve lower write amplification and latency. cLSM~\cite{golan2015scaling} and RocksDB~\cite{rocksdb} use non-blocking synchronization to increase parallelism. We take a different yet complimentary approach to LSM optimization by incorporating models as auxiliary index structures to improve lookup latency, but each of the others are orthogonal and compatible to our core design.

\vspace{0.1in}
\noindent
\textbf{Model choices.} Duvignau et al.~\cite{duvignau2018piecewise} compare a variety of piecewise linear regression algorithms. Greedy-PLR, which we utilize, is a good choice to realize fast lookups, low learning time, and small memory overheads. Neural networks are also widely used to approximate data distributions, especially datasets with complex non-linear structures~\cite{lathuiliere2019comprehensive}. However, theoretical analysis~\cite{livni2014computational} and experiments~\cite{shi2016benchmarking} show that training a complex neural network can be prohibitively expensive. Similar to Greedy-PLR, recent work proposes a one-pass learning algorithm based on splines~\cite{kipf2020radixspline} and identifies that such an algorithm could be useful for learning sorted data in LSMs; we leave their exploration within LSMs for future work.

%% file: fig-error.tex
\begin{figure}[t!] 
\centering
\captionsetup[subfigure]{labelformat=empty, justification=centering}
\subfloat[(a) Error-bound tradeoff]
{
\begin{minipage}{0.25\textwidth}\
 \centering
\includegraphics[scale=0.75]{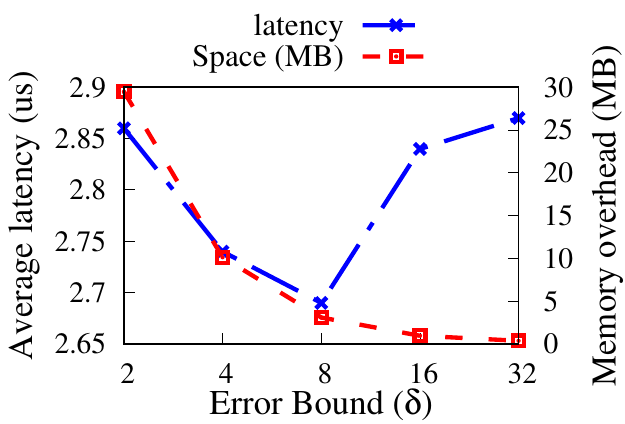}
\end{minipage}%
}
\subfloat[(b) Space overheads]
{
\begin{minipage}{0.22\textwidth}\
 \centering
\input{tbl-space}
\end{minipage}%
}

\vspace{-0.03in}
\mycaption{fig-error}{Error-bound Tradeoffs and Space Overheads}{(a) shows how the PLR error bound affects lookup latency and memory overheads; (b) shows the space overheads for different datasets.}

\end{figure}

%% file: tbl-space.tex
\setlength{\tempa}{\tabcolsep}
\newcolumntype{C}[1]{>{\centering\arraybackslash}p{#1}}
\newcolumntype{M}[1]{>{\centering\arraybackslash}m{#1}}
\setlength{\tabcolsep}{1.1pt}

{\footnotesize
\begin{tabular} {M{1cm}|M{1cm}|M{1cm}}
\multirow{2}{1.5cm}{Dataset} & \multicolumn{2}{c}{Space Overheads} \\\cline{2-3}
									 & MB & \% \\\hline
Linear	&  0.02	& 0.0 \\
Seg1\%	&  15.38	& 0.21 \\
Seg10\%	&  153.6	& 2.05 \\
Normal	&  16.94	& 0.23 \\
AR	&  3.09	& 0.08 \\
OSM	&  7.08	& 0.26 \\
\end{tabular}  
}

\setlength{\tabcolsep}{\tempa}

%% file: conc.tex
\section{Conclusion}
\label{sec-conc}

In this paper, we examine if learned indexes are suitable for write-optimized log-structured merge (LSM) trees. Through in-depth measurements and analysis, we derive a set of guidelines to integrate learned indexes into LSMs. Using these guidelines, we design and build \sysname, a learned-index implementation for a highly-optimized LSM system. We experimentally demonstrate that \sysname\ offers significantly faster lookups for a range of workloads and datasets.

\camera{\sysname\ is an initial work on integrating learned indexes into an LSM-based storage system. More detailed studies, such as more sophisticated cost-benefit analysis, general string support, and different model choices, could be promising for future work. In addition, we believe that \sysname's learning approach may work well in other write-optimized data structures such as the $B^{\epsilon}$-tree~\cite{bepsilon} and could be an interesting avenue for future work. While our work takes initial steps towards integrating learning into production-quality systems, more studies and experience are needed to understand the true utility of learning approaches.}

\if 0 -- old

\camera{\sysname\ is an initial work on integrating Learned Indexes into LSM-tree structure. More detailed studies, such as more sophisticated cost-benefit analysis, general string support, and different model choices, could be promising for future work. In addition, we expect the \sysname\ learning approach may work well in other data structures such as the $B^{\epsilon}$-tree~\cite{bepsilon}. More experience with integration of learned approaches into production-quality systems is needed to understand the true utility of these approaches.}

\fi

%% file: ack.tex
\vspace{0.1in}
\section*{Acknowledgements}

\camera{We thank Alexandra Fedorova (our shepherd) and the anonymous reviewers of OSDI '20 for their insightful comments and suggestions. We thank the members of ADSL for their excellent feedback. We also thank CloudLab~\cite{RicciEide:login14} for providing a great environment to run our experiments and reproduce our results during artifact evaluation. This material was supported by funding from NSF grants CNS-1421033, CNS-1763810 and CNS-1838733, Intel, Microsoft, Seagate, and VMware. Aishwarya Ganesan is supported by a Facebook fellowship. Any opinions, findings, and conclusions or recommendations expressed in this material are those of the authors and may not reflect the views of NSF or any other institutions.}

%% file: bib.tex
{
   \bibliographystyle{plain}
   \bibliography{all-defs,all,ram,urls,yien,all-confs}
}